\documentclass[referee,a4paper,12pt,traditabstract]{swsc} 

\usepackage{graphicx}
\usepackage{txfonts}
\usepackage{subfigure}
\usepackage{epstopdf}
\usepackage[authoryear,round]{natbib}
\usepackage[backref]{hyperref}
\usepackage{url}
\usepackage{pdfpages}
\usepackage[none]{hyphenat}
\hypersetup{colorlinks=true,citecolor=cyan,urlcolor=cyan,linkcolor=blue}
\usepackage[]{xcolor}
\definecolor{CASIIlightindago}{RGB}{60,100,120}
\definecolor{CASIIdarkyellow}{RGB}{188,161,54}
\definecolor{CASIIdarkgreen}{RGB}{75,85,50}
\definecolor{CASIIaliceblue}{RGB}{16,120,150}
\begin{document}


   \title{Application Usability Levels: }

   \subtitle{A Framework for Tracking Project Product Progress}
   
   \titlerunning{Application Usability Levels}

   \authorrunning{Halford et al}

\author{
Alexa J. Halford\inst{1}, 
Adam C. Kellerman\inst{2},
Katherine Garcia-Sage\inst{3,4},
Jeffrey Klenzing\inst{4}, 
Brett A. Carter\inst{5},
Ryan M. McGranaghan\inst{6,7},
Timothy Guild\inst{1},
Consuelo Cid \inst{8},
Carl J. Henney\inst{9},
Natalia Yu. Ganushkina\inst{10, 11},
Angeline G. Burrell\inst{12},
Mike Terkildsen\inst{13},
Daniel T. Welling\inst{14, 10},
Sophie A. Murray\inst{15},
K.~D. Leka\inst{16},
James P. McCollough\inst{9},
Barbara J. Thompson\inst{4}, 
Antti Pulkkinen\inst{4},
Shing F. Fung\inst{4},
Suzy Bingham\inst{17},
Mario M. Bisi\inst{18},
Michael W. Liemohn\inst{10},
Brian M. Walsh\inst{19}, \and
Steven K. Morley\inst{20}}
%

\institute{Space Sciences Department, Aerospace Corporation, Chantilly, Virginia, USA.  \email{\href{mailto:alexa.j.halford@aero.org}{Alexa.J.Halford@aero.org}} \and
Department of Earth, Planetary, and Space Sciences, University of California, Los Angeles, USA. \and
Catholic University of America, Washington D.C., USA.\and
NASA GSFC, Heliophysics Science Division Greenbelt, MD 20771, USA. \and
School of Science, RMIT University, Melbourne, Australia. \and
NASA Jet Propulsion Laboratory, California Institute of Technology, Pasadena, California, USA. \and
University Corporation for Atmospheric Research, Boulder, Colorado, USA. \and
Departamento de F\'{i}sica y Matem\'{a}ticas, Universidad de Alcal\'{a}, Alcal\'{a} de Henares (Madrid), Spain. \and
Space Vehicles Directorate, Air Force Research Laboratory, Kirtland AFB, New Mexico, USA. \and
University of Michigan Climate and Space department, Ann Arbor, MI, USA. \and
Finnish Meteorological Institute, Helsinki, Finland. \and
Space Science Division, U.S. Naval Research Laboratory, Washington, DC 20375, USA. \and
Space Weather Services, Bureau of Meteorology, Sydney, Australia. \and
Physics Department, University of Texas at Arlington, Arlington, TX, USA. \and
School of Physics, Trinity College Dublin, Ireland. \and
NorthWest Research Associates, Boulder, Colorado, USA. \and
Met Office, Fitzroy Road, Exeter, Devon, EX1 3PB, UK. \and
RAL Space, Science \& Technology Facilities Council - Rutherford Appleton Laboratory, Harwell Campus, Oxfordshire, OX11 0QX, UK. \and
Center for Space Physics, Boston University, Boston, USA. \and
Space Science and Applications, Los Alamos National Laboratory, Los Alamos, NM, USA.}


 
  \abstract
   {The space physics community continues to grow and become both more interdisciplinary and more intertwined with commercial and government operations. This has created a need for a framework to easily identify what projects can be used for specific applications and how close the tool is to routine autonomous or on-demand implementation and operation. We propose the Application Usability Level (AUL) framework and publicizing AULs to help the community quantify the progress of successful applications, metrics, and validation efforts. This framework will also aid the scientific community by supplying the type of information needed to build off of previously published work and publicizing the applications and requirements needed by the user communities. In this paper, we define the AUL framework, outline the milestones required for progression to higher AULs, and provide example projects utilizing the AUL framework. This work has been completed as part of the activities of the Assessment of Understanding and Quantifying Progress working group which is part of the International Forum for Space Weather Capabilities Assessment.}           


   \maketitle

\section{Introduction}

As a field, space physics has quickly evolved beyond science inquiries and pure research. We are currently at the point where new opportunities and a need for interdisciplinary and applied space weather research have notably increased. As such, research-to-research and research-to-operations communication frameworks have become important tools. These tools expedite both multidisciplinary research and the transition of research tools to applications. It is important that, as a community, we are able to identify which research applications are ready to be transitioned into technology and provide useful information for users. It is important that the scientific community is able to clearly communicate the progress of the transition process. It is equally important to provide a measure of the usability of a research project to a user-defined application. To effectively accomplish this task, researchers need clear communication of a user's requirements, needs, and metrics for successful use of an application. In this paper, we introduce a new framework to aid in communication and collaboration of space physics research applications.

\subsection{Previous Tracking Frameworks}
Our new framework was developed to address needs not met by existing tracking frameworks.  The most well-known example of such a tracking framework is the Technology Readiness Level (TRL) system, which categorizes the maturity of a particular technology and its use for instrumentation \cite[e.g.,][and references therein; European Space Agency (ESA) TRL definitions can be found at http://sci.esa.int/sci-ft/50124-technology-readiness-level/]{mankins1995technology, MANKINS2009, Azizian2011, olechowski2015}.
The clear, consistent definitions of the TRLs allow the readiness of an instrument for a specific use to be determined independently and in comparison to currently available options.

NASA's Earth Sciences division's applied science program employs the Application Readiness Level (ARL) framework. The ARL framework is used to communicate how ``ready'' a given model or data analysis effort is for a particular utilization and industry partner. The framework's focus on research and development involves the industry partner at the project's start \cite[see Figure 1, or https://www.nasa.gov/sites/default/files/files/ExpandedARLDefinitions4813.pdf]{Pulkkinen2017}. This includes forming the application requirements around the user needs. This framework has aided in the identification of obstacles. It is also used to assess programmatic health by comparing a project's progression through the framework against the distribution of funding across projects at different levels(e.g., see the Earth Sciences Division, Applied Science Programs Annual Reports: https://appliedsciences.nasa.gov/library-page).  

Both ARLs and TRLs use single-digit level identifiers. They are used to communicate the advancement of applied products to the scientific, engineering, and funding communities. Though the two frameworks are designed for different types of products, the identifiers communicate similar information.  The level identifiers enable researchers, funding agencies, and users to easily interact with each other and communicate progress towards the routine usage of these products.

\subsection{A New Tracking Framework for Heliophysics}
The existing frameworks each focus on tracking a particular type of product, and so do not fully meet the needs of the heliophysics community. Space physics products include observational data, derived indices, modeled outputs, and more. These products are often used together for different purposes. Each user will have different requirements for the application in terms of the type of product, robustness, and accuracy. 

The unique needs of the space weather community led to the modification of existing research-to-application communication frameworks to create the Application Usability Level (AUL) framework.  Applying AULs to model and data analysis efforts can benefit space physics research. These benefits include improving access to collaborators, project transparency, and communication of project results. As the requirements and user interests for each application are unique, the AUL framework uses specifically-tuned metrics.  For instance, a research user interested in upper atmospheric coupling may want to know the flux and characteristic energy of precipitating electrons. Similarly, a satellite industry partner may want to predict satellite drag during a geomagnetic storm. A single research project may be able to provide both users with the products they need. However, the different outputs will require different metrics, implementation strategies, and time frames for implementation. Since the AUL framework is highly adaptable, it can help a single research project meet and track both of these user needs. 

The AUL framework can bolster communication between researchers, users, funding bodies, and stakeholders. Using a standard framework provides a clear path for users and researchers to follow. This improves efficiency assuring that all components from the researchers' project to the user needs are considered. It enables communication about a proposal's development status, requirements for further progress, and achievable goals. Improved communication leads to better-targeted funding opportunities and proposals. The AUL framework can simplify comparisons of different projects working towards a specific application. This enables operational and funding services to select the most appropriate proposal for their requirements.  It can also highlight gaps in knowledge, data, and technology, to aid the characterization of needs for new missions, instruments, and research or model development proposal calls.

In this article, we introduce the AUL framework, a new heliophysics-focused research-to-application framework for communicating the usability and readiness of a product to its user.  The AUL framework is constructed similarly to the existing TRL and ARL frameworks.  The framework terminology is described in section~\ref{Def}. The details of the AUL framework are provided in section~\ref{AUL}.  Section~\ref{examples} provides examples of AULs at different usability levels.  The potential impact of using AULs is described in section~\ref{summary}. Full examples and tools to help aid in the adaptation of AULs are provided in the appendix and supplemental information. 

\section{Framework Terminology for Targeted Research}\label{Def}

The AUL framework can be applied to any project where an expected outcome is the ongoing use of a product by another party (targeted research). In a very general sense, the framework can be applied as described in subsection \ref{AUL_par}.  This paragraph contains italicized words that we have identified and defined in subsection \ref{AUL_term}.  These definitions are included to avoid the confusion commonly encountered in interdisciplinary, multidisciplinary, and transdisciplinary projects. One of the strengths of AULs is its ability to enable communication between different groups. This includes researchers in the same field (unidisciplinary or interdisciplinary), scientists across disciplines (multidisciplinary or transdisciplinary), and with industry partners (transdisciplinary or applied).

\subsection{AUL Framework General Use Example}\label{AUL_par}
We, a group of researchers, have a {\it project}. We believe that there is an {\it application} that it can be used for and have determined that we will use the {\it AUL} framework and its {\it phases} to communicate the progress of this {\it project} towards a specific {\it application} to the specified {\it user}, as well as to the scientific community. First, we will identify and reach out to a potential {\it user} who might be interested in routinely using the {\it project}'s {\it product}. We then determine if the {\it project} is {\it viable} based on the {\it user}'s {\it requirements}. If it is, then we continue by defining {\it metrics} for {\it verifying} the {\it viability} and {\it feasibility} for the specific {\it application} with the {\it  user}'s {\it requirements} in mind. If the project is not deemed {\it viable} or {\it feasible}, then the current {\it project} should be re-examined and potentially held off on until it is deemed both {\it viable} and {\it feasible}.  As the {\it project} continues, dissemination of the progress, {\it metrics}, and {\it validation} efforts should be reported to both the {\it user} and the relevant community.  Once the {\it product} is {\it validated} and demonstrated to work within the {\it relevant context}, it is {\it transitioned} over to regular on-demand use. {\it Validation}  and {\it verification} efforts continue, now focusing on  sustained usage in the {\it operating environment}. Each step in this development effort towards application readiness is given an {\it AUL} level number. For every step in this process, an {\it AUL} number is designated to denote and communicate the current state of the {\it project}. 

\subsection{AUL Terminology} \label{AUL_term}
\begin{description}

\item[Application -] A specific use for a project, such as a data product from a mission, a service such as satellite hardware anomaly assessments, or a forecast of a specific quantity from a numerical model.  Each application has its own unique requirements and metrics for validation. For instance, an application may be forecasts of surface charging events that have an 85\% success rate and less than a 5\% rate of false alarms. 

\item[Applied Research -] Research pursued with a focus on providing a practical application or new technology.\\

\item[AUL -] The Application Usability Levels (AULs) are the scale that tracks the progress of work on a given project for a specific application, as summarized in Figure \ref{Fig_AUL} and Table \ref{Tab_AUL}.  More details about the three phases and nine levels of the AUL framework are found in Section \ref{AUL}. \\

\begin{figure}
\centering\includegraphics[width=0.9\linewidth]{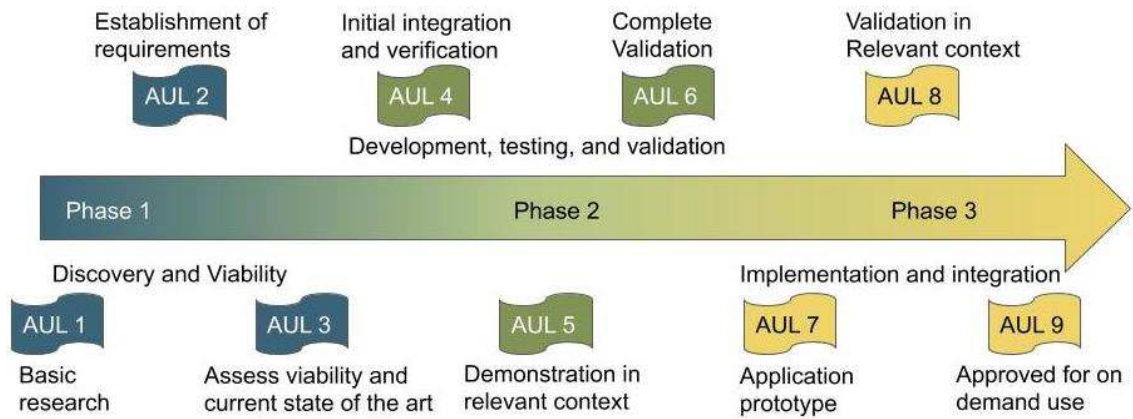}
\caption{Application Usability Level (AUL) diagram. The progress of a project towards an application moves from AUL 1 to AUL 9, passing through three main phases: discovery, development, and implementation. Each of these are described in the main text. }\label{Fig_AUL}

\end{figure}

\item[Feasibility -] The ability to achieve success with the available resources. \\ 

\item[Metric -] A quantitative measure of project or application performance. When applied to project progress, this constitutes measures that define whether the project meets its goals and milestones. When applied to applications, metrics consist of quantities appropriate for measuring performance, such as accuracy, bias, or skill score. \\

\item[Operations to Research (O2R) -] Through the process of transitioning targeted or applied research to operations, new phenomena or discoveries can be found and inform subsequent research projects. This part of the feedback process is referred to as Operations to Research (O2R)\\

\item[Operational Environment -] The conditions in which the application will be used. For example, a geophysical research model that will be delivered to the Community Coordinated Modeling Center (CCMC) for on-demand runs will define their operation environment as the computer system used by CCMC.\\

\item[Phases -] AULs are grouped into three phases: 1) discovery and viability; 2) development, testing, and validation; and 3) usability, final validation, and implementation. Each of these phases is discussed in more detail in the following sections and summarized in Figure \ref{Fig_AUL} and Table \ref{Tab_AUL}. \\

\item[Product -] A project outcome that is routinely used to enhance the decision-making process of a user or provide input to another research project or application.\\

\item[Project -] A research or development initiative designed to make progress towards a single application. Examples of projects include the development or modification of models, new uses for available data, using a data product to improve decision making, and using current knowledge to develop future projects.\\

\item[Research to Research (R2R) -] A targeted research project where both the 'researcher' and 'user' are researchers who may be in the same sub-field, or in completely different disciplines. \\

\item[Research to Operation (R2O) -] A targeted or applied research project which takes a research application and transitions it into the operational environment.  \\

\item[Requirements -] The set of necessary conditions outlined by the user, which may include metrics, time frames, and operational environments that the project must meet for the resulting application to be considered successful. \\

\item[Relevant context -] The environment in which the project must be validated and verified (e.g., during geomagnetic storm periods or in interplanetary space). \\ 

\item[Targeted Research -] Investigations pursued with a specific objective. \\ 

\item[Transition -] The process or set of activities that take a product or service from a testing environment and move it to an operational environment. \\

\item[User -] The anticipated person or group who will make use of or operate the project's application. This may be another researcher, broker, or industry partner. Other common terms for user, appropriate for different fields, include `end user', `forecaster', `customer', or even `another collaborator'. \\

\item[Validation -] The determination of the skill of the project's outputs, quantified by identified metrics for the defined operating environment and relevant context.\\

\item[Verification -] The determination that the product conforms with the project requirements, as described in the relevant design documents.\\

\item[Viability -] The project's value or level of return for the researchers and users. 

\end{description}

\noindent Terms may have different definitions within different communities. For example, the terms {\it validation} and {\it verification} can be particularly confusing. Their definitions differ between the operational and modeling communities. The operational community typically uses {\it verification} to mean that a system meets end-user requirements and uses {\it validation} to mean that the system is right for the project. The modeling community typically uses {\it verification} to mean that the code operates correctly and uses {\it validation} to mean that the results are accurate. For this paper, we have settled on the listed definitions based on commonalities between the two communities. Both identify {\it verification} as meeting basic requirements and {\it validation} as the appropriateness of the result within a given environment.

\section{The Application Usability Level (AUL) Framework}\label{AUL}

The new AUL framework benefits research by providing a structured approach for tracking the progress of a project towards an application. Once the needs of a specific application have been defined, the same metrics may be used by the community to assess the progress of several projects towards the same application. This allows for easy comparison of projects and provides insight into a project's progress. In this section, we will define each of the three phases in the framework, the individual levels that make up each phase and step through the necessary milestones to achieve a given AUL. A summary figure of the AUL levels and phases can be seen in Figure \ref{Fig_AUL} and is outlined in Table \ref{Tab_AUL}. A checklist for the three phases is provided in Appendix \ref{checklist}. More resources can be found through our team website at https://ccmc.gsfc.nasa.gov/assessment/topics/trackprogress.php. 

\begin{table}
\caption{A brief description of the AUL phases and levels}
\centering
\begin{tabular}{llll}\hline
Phase & Phase definition & AUL & Level description \\
 \hline
 & & \textcolor{CASIIaliceblue}{1}& \textcolor{CASIIaliceblue}{Basic research}\\
 \textcolor{CASIIaliceblue}{{\bf Phase 1} }& \textcolor{CASIIaliceblue}{{\bf Discovery and Viability }} & \textcolor{CASIIaliceblue}{2} & \textcolor{CASIIaliceblue}{Establishment of users and their requirements}\\ 
& & \textcolor{CASIIaliceblue}{3} &\textcolor{CASIIaliceblue}{Assess viability and current state of the art} \\
 \hline
 \hline
 & & \textcolor{CASIIdarkgreen}{4}& \textcolor{CASIIdarkgreen}{Initial integration and verification}\\
 \textcolor{CASIIdarkgreen}{{\bf Phase 2} }& \textcolor{CASIIdarkgreen}{{\bf Development, Testing,  }} & \textcolor{CASIIdarkgreen}{5} & \textcolor{CASIIdarkgreen}{Demonstration in the relevant context}\\ 
& \textcolor{CASIIdarkgreen}{{\bf and Validation}}& \textcolor{CASIIdarkgreen}{6} &\textcolor{CASIIdarkgreen}{Completed validation} \\
 \hline 
 \hline
 & & \textcolor{CASIIdarkyellow}{{7}}& \textcolor{CASIIdarkyellow}{{Application prototype}}\\
 \textcolor{CASIIdarkyellow}{{{\bf Phase 3}}}& \textcolor{CASIIdarkyellow}{{{\bf Implementation and Integration}}} & \textcolor{CASIIdarkyellow}{{8}} & \textcolor{CASIIdarkyellow}{{Validation in relevant context}}\\ 
& \textcolor{CASIIdarkyellow}{{{\bf into Operation}}} & \textcolor{CASIIdarkyellow}{{9}} &\textcolor{CASIIdarkyellow}{{Approved for on-demand use}} \\

\label{Tab_AUL}
\end{tabular}
\end{table}

\subsection{\textcolor{CASIIaliceblue}{Phase I: Discovery and Viability}}
Phase I is where fundamental research meets applied science. Not all research will or should progress beyond the very first AUL. However, if a potential user is identified (whether they are a fellow researcher or an industry partner), then the steps in this phase will determine whether the project should progress to phase II. Phase I provides researchers with justifiable confidence that their work is leading to a product for a specific application that will provide the user with the information they need. The first step in the progression of any project towards an application requires contacting a potential user to begin forming a partnership. The user's needs must be established through effective and clear communication about the requirements and metrics of the application. The researchers must determine if the project will be viable and feasible to satisfy the requirements for that specific application. It should be determined whether the current project represents the current standard or an improvement upon the state-of-the-art for that specific application. 

\subsubsection{AUL 1: Basic Research}
This level is where the basic scientific concepts and projects are created and potential applications are identified. A project is considered to have an AUL 1 if the following milestones are achieved: \\

\noindent {\bf Milestones:}
\renewcommand{\labelenumi}{\alph{enumi})}
\begin{enumerate}
\item  Basic research is documented and disseminated for the project, so that the usability may be assessed by way of the AUL method. 
\item  Ideas for how the project output(s) may enhance decision making or be applied to an end user application are generated.
\item  Potential interested users are identified, but not necessarily contacted.  This could occur, for example, through a literature search, conference attendance, or workshop participation.
\end{enumerate}

\subsubsection{AUL 2: Establishment of users and their requirements for a specific application}
In this level, the application and project concept is formalized. An interested user is contacted, and their needs for a specific application are identified. This includes establishing requirements and defining metrics for measuring success. It should be noted that at this level, it is not necessary to show that the current research endeavors will result in the successful production of the identified application. A project is considered to be at AUL 2 if the following milestones are achieved: \\

\noindent {\bf Milestones:}
\begin{enumerate}
\item Decide on the user(s), contact the user(s), and establish a reliable channel of communication that is used at a suitable frequency.
\item Formalization of the application and project concept.
\item Identification and formalization of the requirements and metrics necessary for successful application of the project for the user's needs. 
\end{enumerate}

\subsubsection{AUL 3: Assess viability of concept and current state of the art }
To reach AUL 3, the feasibility and viability of achieving success for a specific application as defined in AUL 2, must be assessed by both the users and researchers. Building upon a proof-of-concept study, the requirements for project improvement and chosen metrics are re-examined and updated as needed. A demonstration that the project represents the state-of-the-art, an improvement, or value-add to the state-of-the-art is completed. A project is considered to have an AUL 3 if the following milestones are achieved: \\

\noindent {\bf Milestones:}
\begin{enumerate}
\item Documentation and dissemination of the project's expected advancements from the current state-of-the-art used towards the identified application along with the proposed metrics for the specified application. 
\item Perform the initial analysis of the individual project components, to determine the viability and feasibility of the entire project.
\item Complete a detailed characterization of the baseline performance and limitations with respect to the application.
\item Determine the viability and feasibility of the proposed project towards improving upon the state of the art for the identified application. If the project is deemed not viable or feasible, the project is put on hold until the identified roadblocks are removed.
\end{enumerate}

\subsection{\textcolor{CASIIdarkgreen}{Phase II: Development testing and validation}}
In Phase I, the current state of the art is identified,  basic research into current limitations and expected areas for improvements are completed, initial communication with the end user is established, and a proof-of-concept and show of viability  is made. Phase II  focuses on finalizing development of the new state-of-the-art project integrating the resulting tools into the identified applications, demonstrating the feasibility of the new product and validating the new system. 

\subsubsection{AUL 4: Initial integration and verification}
In this level, the basic prototype is completed and initial integration into the user application is started. To achieve AUL 4, it must be verified that all components work together. In addition, a project is considered to have AUL 4 if the following milestones are also achieved: \\

\noindent {\bf Milestones:}
\begin{enumerate}
\item Integration of the individual components into the application.
\item Organizational challenges and human process issues (if applicable) are identified and managed. 
\end{enumerate}

\subsubsection{AUL 5: Demonstration in the relevant context}
In this level, the viability of the project is determined for the specified relevant context (e.g., storm, substorm, or quiet time conditions). A project is considered to have AUL 5 if the following milestones are achieved: \\

\noindent {\bf Milestones:}
\begin{enumerate}
\item The project team must articulate and disseminate the viability  for the improvement upon the state of the art.
\item Application components integrated into a functioning application system for use during the given relevant context parameters. 
\end{enumerate}

\subsubsection{AUL 6: Complete validation}
While in AUL 5 the potential is articulated, in AUL 6 the potential is fully demonstrated, and this is stated as a major increase in the applications usability and ability to become the new standard for the user. Any application components already deployed in the  user's operating environment are tested in their operational and/or decision making context. A project is considered to have AUL 6 if the following milestones are achieved:\\

\noindent {\bf Milestones:}
\begin{enumerate}
\item Prototype application system beta-tested in a simulated operational environment.
\item Projected improvements in performance of the state-of-the-art and/or decision making activity demonstrated in simulated operational environment.  
\item Documentation and dissemination of the specific application and associated metrics and the projects progress towards this application.
\end{enumerate}

\subsection{\textcolor{CASIIdarkyellow}{Phase III: Implementation and Integration into operational status}}
While Phase I and II focused on the development and initial validation of the new model/data analysis effort for a specific application, Phase III is where it becomes handed off and fully integrated into the end user's application. This also includes new validation efforts to determine how well the new model/data analysis effort performs in a ``real world" setting. Validation and continued use in an operational environment drive the discovery of new scientific questions, problems, and new applications. Although this is the final phase for the current application with its specific requirements and metrics, the search for the next new and improved application continues. 

\subsubsection{AUL 7: Application prototype}
All portions of the new project are integrated into the user's application and the functionality has been established. A project is considered to have AUL 7 if the following milestones are achieved: \\

\noindent {\bf Milestones:}
\begin{enumerate}
\item The system must be fully integrated into the operational environment specified by the user. 
\item The system's functionality is tested and demonstrated in the user's specified relevant context. 
\item Project team must demonstrate the functionality of the new system for the user's application and disseminate the results.
\end{enumerate}

\subsubsection{AUL 8: Validation in relevant ``real world'' environment}

At AUL 8, the new project is fully integrated into the user application system and is initially validated by the user. The application is proven to work in its final form in the relevant context and operational environment either meeting or surpassing the initially identified requirements and metrics. In addition, user documentation including verification and validation/metric results, any limitations of the new project, training documentation, and maintenance documentation are completed. Ideas for future developments are documented. A project is considered to have AUL 8 if the following milestones are achieved: \\

\noindent {\bf Milestones:}
\begin{enumerate}
\item The user must approve the addition of the new project to their operational application system. 
\item Finalized application system tested, proven operational, and shown to operate within the specified requirements and metrics. 
\item Applications qualified and approved by the user. 
\item User documentation and training completed. 
\end{enumerate}

\subsubsection{AUL 9: Approved for on-demand use towards stated application}
At AUL 9, the project is the new state of the art and has been proven to work in a sustained manner. Continued validation efforts, completed by the user(s) and likely in concert with the researcher(s), are performed for the project's sustained use in the operational environment. A project is considered to have AUL 9 if the following milestones are achieved: \\

\noindent {\bf Milestones:}
\begin{enumerate}
\item Sustained and repeated use of the application by the specified users.
\item The continued validation of the project in the operational environment. 
\item Dissemination of the validation efforts, metrics, and new state of the art project to the relevant community for the specific application. 
\end{enumerate}

\begin{figure}
\centering\includegraphics[width =.6\linewidth]{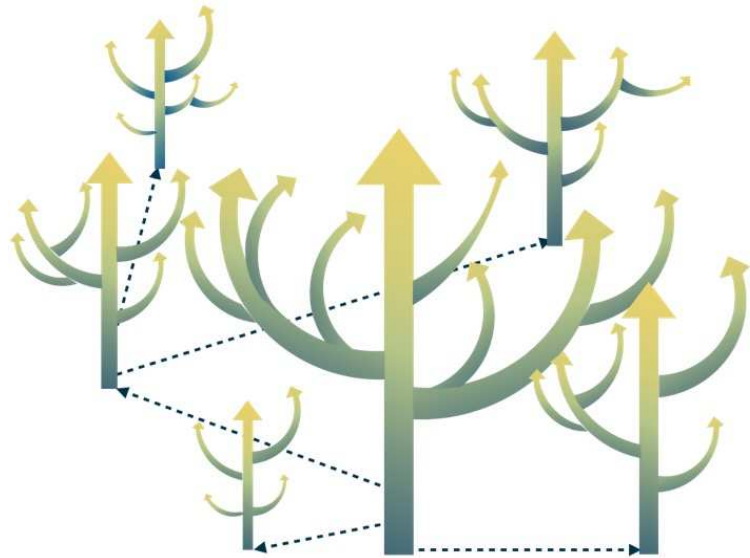}
  \caption{Each AUL can spawn many branches through working with users and whole new trees which may remain connected through their common roots, the same basic research (like an aspen grove) as suggested in Sections \ref{Jeff} and \ref{Brett}. These new applications may be identified at any stage through the process and will have their own users, requirements, and metrics and will progress through the AUL framework at their own pace.}
  \label{fig_branch}
\end{figure}

\begin{figure}
  \centering \includegraphics[width=.7\linewidth]{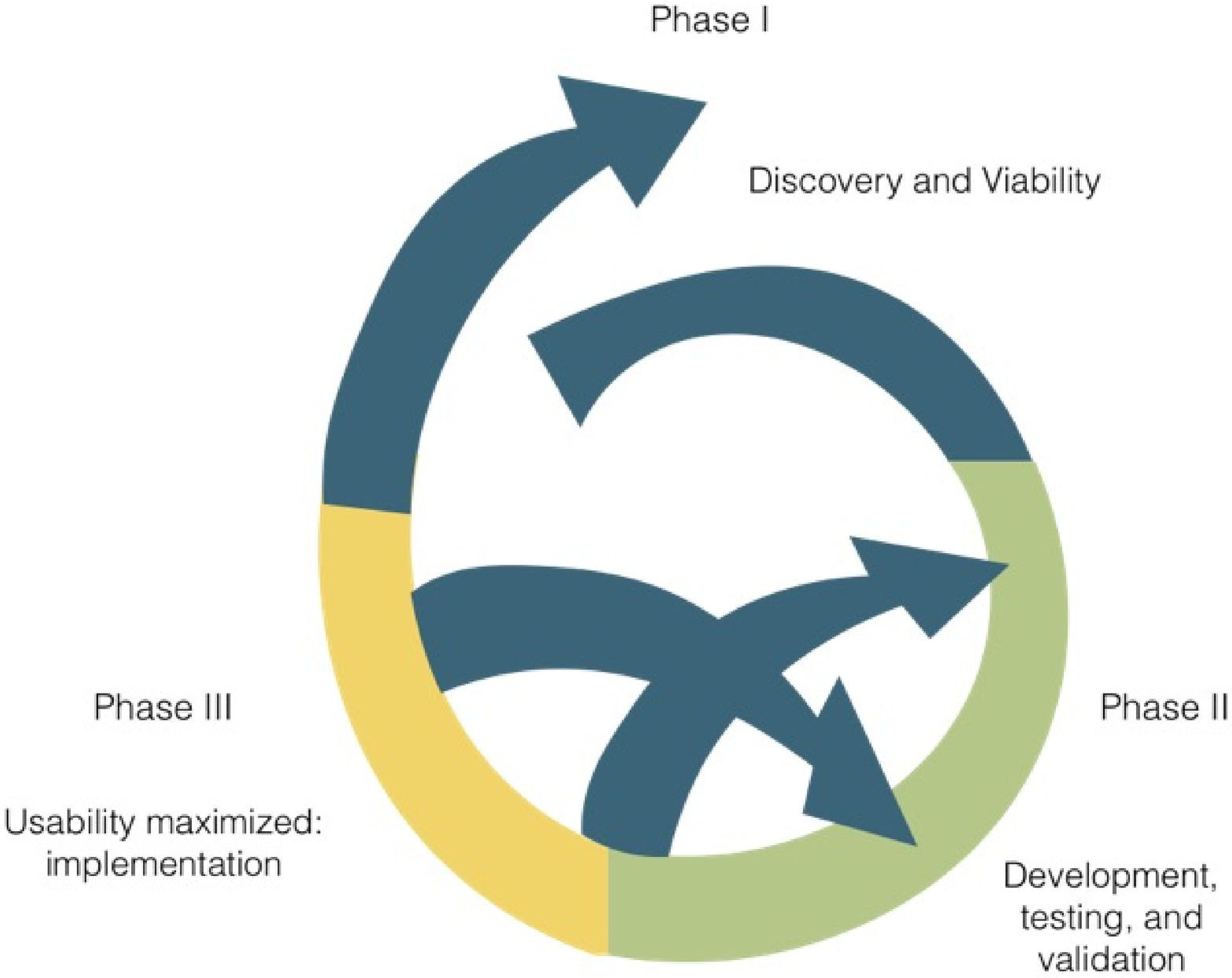}
  \caption{While working through an AUL, roadblocks can appear which will temporarily lower the AUL of the project, similar to how a change in an instrument project can lower the TRL of a hardware project. An example of this would be a component of a large model may have changed and need to be validate, or a change to the processing of the data which is an input to the project. Anecdotal experience has shown how reaching an AUL 9 does not mean the end of a project. By the time an application reaches AUL 9, new requirements are often defined and a new application and project are created. This may be as simple as the same user has identified the need of new requirements and metrics for a new (perhaps an improvement on the same, but now identified as a new) application, leading to a spiraling of the AULs, as suggested and demonstrated in Sections \ref{Adam} and \ref{Cid}}
  \label{fig_AUL_spiral}
\end{figure}

\subsection{The next application}
The AUL outline and process shows the path and progress from research to application. It does not as explicitly show the feedback from application or operations to research. As with all research, there are always new questions and better tools and methods which become available and allow for improvements such as, better forecasting, smaller error bars. Once a project has reached AUL 9, there are often new areas for improvements which have been identified, new science questions which were uncovered during the process, and of course, new potential applications realized through working with the users. This framework is less of a line and more of a set of branching trees as shown in Figure \ref{fig_branch} or a spiral as shown in Figure \ref{fig_AUL_spiral}.  In sections \ref{examples} and \ref{append}, the AUL framework is applied to cutting edge research and R2O processes. In many of these examples, refinements of the metric criteria and completely new applications or users are identified. Thus, with each new application (along with specific metrics and operating criteria), the model or data product begins a new AUL branch.

\subsection{Dissemination of Results}

Throughout the AUL framework, there are 6 milestones that require the dissemination of results. AUL 1 requires that the basic science has been documented. AUL 3 requires that the expected advancements are shared while AUL 5 requires documentation of the viability for improvement upon the state of the art. All three of these milestones help determine the viability and feasibility of the proposed project to address the needs of the user. The finalized metrics and requirements of the specific application are documented in AUL 6, and in AUL 7 the new functionality of the project in the user environment is shown. Continued validation, and how the project is the new operational state of the art is documented in AUL 9. \\

\noindent{\bf Peer Reviewed Papers:}
Although it may depend on who the researchers and users are, one natural method for dissemination of the project's progress is through peer-reviewed papers. For many projects, this would be the ideal method for completing the relevant milestones in AULs 1, 6, and 9. This has two primary benefits; 1) the assessment of the new AUL standing is reviewed by an independent assessor and 2) it advertises the application and the advancements of the project. In order to help facilitate the adaption of instrument-like or ``AUL papers", we have provided both \LaTeX and Microsoft word templates on our team website at the CCMC and as of this publication can be found at https://ccmc.gsfc.nasa.gov/assessment/topics/trackprogress.php. We expect that an AUL paper would be a concise piece, outlining the application, the researcher, the user, and the requirements along with the relevant advancements through the milestones. It should also include references to previous AUL papers or other research actives (papers, conference talks, ect.) to show previous progress to the current AUL as well as any new milestones met. These templates may also guide the introduction of AULs into more traditional paper formats.

The above paragraph assumes that the researcher would be writing the AUL papers, however, it has been proposed that the user may also be interested in writing such papers. One of the current hurdles in completing targeted, transdisciplinary, or applied research is identifying who may have the required research and who may benefit from the research. It has been suggested at workshops on AULs that users may be interested in writing AUL 1 papers as a call for help finding interested researchers to address their needs. \\

\noindent{\bf Conference Talks and Posters:}
Presenting results and advancements at conferences is another method for disseminating results and advancements through the AUL framework. For many projects, this would be the ideal method for completing the relevant milestones in AULs 3, 5, and 7. Often it is expected that these advancements would be shared within the appropriate science sessions. However, the Assessment of Understanding and Quantifying Progress working group has convened sessions at multiple conferences and workshops including the Geospace Environment Modeling (GEM) workshop and the American Geophysical Union Fall meeting, that have focused on sharing examples of AUL progressions, and discussions of research to operations and transdisciplinary research. We encourage and support the idea of more sessions at conferences focused on reporting on new applications, advancements of projects, and lessons learned.\\

\noindent{\bf Websites and Online Documentation: }
Peer-reviewed papers and conference presentations are not the only methods for disseminating results, although they are perhaps the most familiar to researchers.  The Assessment of Understanding and Quantifying Progress working group which is part of the International Forum for Space Weather Capabilities Assessment has been developing a website for the dissemination and tracking of applications and projects. The website will host applications with the specified end users and their requirements and the projects working towards the specific application. This will also allow for new researchers to find applications and their requirements which may be applicable for their research. It will also enable users to find applications relevant to their need and researchers who may be able to address their application needs.


\subsection{Best Practices}
The above framework was purposefully written in a general format so that it can be adapted to the needs of the researcher, the user, and the specific project. However, many projects will have a set of best practices (e.g., software \cite[e.g.,][]{Burrell2018}, data set production \cite[e.g.,][]{Wilkinson2016}) and below we will discuss a few which will be important for many if not all projects and applications. \\

\noindent{\bf Training:} 
In AUL8, milestone d) training must be completed. This is a very important step in making sure that the user understands the full capabilities of the product, and will use the product properly. It is necessary to train not just on the typical types of events or runs expected but to also train for the less likely or extreme scenarios. This is especially important for applications related to space weather forecasting where training should occur on both real-time data, and the range of events where different decisions would need to be made. \\

\noindent{\bf Determination of availability of data and robustness of the system:}
When looking at the feasibility of a project, it is important to consider the impact of data outages. This can be mitigated by having redundancy in the system. It is important to test the robustness, the likelihood of an outage in order to determine if the project is feasible. \\

\noindent{\bf Version Control:} 
As data, codes, and programming languages evolve, version control is vital.  Version control ensures that stable releases are available, while further testing and development continue. As new data, models, etc. are developed, they will also go through the AUL framework to show their advancement to the application, and validate their usage. \\

\noindent{\bf Continued testing of operational availability:} 
In phase 3, best practices dictate that project validation and the determination of a project's operational availability continue.  \\

\noindent{\bf Standardized formats:} 
There are likely to be either multiple projects working towards a single application (e.g., the CME arrival scoreboard, or for data management FAIR \cite[e.g.,][]{Wilkinson2016}) or multiple projects which use similar data types. In order to make it easy for others to adopt the outputs of the projects, or use the data for inputs, it is important to use data and meta-data standardized practices. There are many groups working towards defining the appropriate set of standardizations including the Information Architecture group through the International Forum for Space Weather Capabilities and Assessment. 

These are just a few of the best practices one needs to keep in mind \cite[for example, software best practices are outlined in][]{Burrell2018}, but they largely fall under the AULs which are focused on determining either the feasibility and viability of the project, or in the transitioning of the project from the researcher to the user. It is important for both the researcher and user communities to continue to evaluate what best practices are needed and include them when completing the relevant milestones.

\section{Summary of Example Projects using AULs in the Appendix}\label{examples}
Here we will present a summary of examples that cover many different aspects of the heliosphere and how the AUL framework can be applied. Extended versions of these examples can be found in the appendix \ref{append}. For each example in the appendix, we'll reflect on how the AUL framework could be used to assess the progress of each project towards the specified application. Many of the projects were initially developed without the AUL framework, however, they all exhibited best practices and hence we can use the AUL framework to measure their progress, and point out how and where it may be useful in each case. The Phase I Example \ref{Jeff}, describes a research area which has identified multiple potential applications but has not yet identified a user. Example \ref{Jeff} also provides a look at how to start the AUL process and targeted research more generally.  The Phase II examples include Example \ref{Ryan} has a research user whereas Example \ref{Brett} which has a  government agency as the user and Example \ref{Tim} has an industry user. Example \ref{Adam} shows how as a project moves through the AUL levels, new applications can be found by a change in the requirements of the user.  Example \ref{Carl} shows that there may be many different user communities interested in the product and application. The Phase III examples include Example \ref{Cid} where one can see how an AUL 9 project often leads to the identification of a new application and thus a new project at AUL 1. And finally, Example \ref{NYG} shows how a project may not be finished through a single funding opportunity, and in this specific example, each phase in the AUL framework was funded through a separate opportunity. 

For the ease of the reader, we have provided a table of the examples, a brief summary of the project, phase, and user, which section they are in, where the longer version of the example can be found (Table \ref{tab_examples}).

\begin{table}[]
\caption{Table of examples given in the paper}
\label{tab_examples}
\begin{tabular}{p{3.75cm}p{0.75cm}p{0.75cm}p{2.5cm}p{3cm}p{1.5cm}}
Example    & Phase & AUL  & Research Sub-field & Primary User &longer example    \\  \hline
\ref{Jeff} Identifying a new application   & 1     & 1    & Ionosphere        & Researcher models &  \ref{append:Jeff} \\  \hline
\ref{Ryan} Application for another researcher     & 2     & 5    & Ionosphere        & The AMIE model &\ref{append:Ryan}   \\  \hline
\ref{Adam} Branching applications   & 1, 2   & 2, 5 & Magnetosphere   & External Business & \ref{append:Adam}    \\  \hline
\ref{Brett} Transitioning a research model to a government user & 2  & 6  & Ionosphere & Australian Bureau of Meteorology & \ref{append:Brett}\\  \hline

\ref{Tim} Validating in an operational environment    & 2     & 6    & Magnetosphere     & Industry/government & \ref{append:Tim}\\  \hline
\ref{Carl} Identifying new transformative research by working with the user & 2 & 7  & Solar   & Government/Air Force & \ref{append:Carl} \\ \hline
\ref{Cid} Identifying new applications   & 1, 3     & 1, 9    & GICs    & Red El\'ectrica de Espa\~na, REE & \\ \hline 
\ref{NYG} Funding applications through the three phases     & 3     & 9    & Magnetosphere     & British Antarctic Survey &\ref{append:NYG} \\     
\end{tabular}
\end{table}

\subsection{Identifying a potential new application to track with the AUL framework: Phase I AUL 1 project}\label{Jeff}
\noindent {\bf  J. Klenzing and A. G. Burrell}

The following is a summary of a phase 1 AUL 1 example from the ionospheric community. This example shows a data project, and how one can start using the AUL framework. A full version of this example can be found in Appendix \ref{append:Jeff} 

The use of average solar extreme ultraviolet (EUV) as a potential driver of ionosphere-thermosphere (I-T) models is an example of an early-phase AUL 1 project.  This radiation heats the thermosphere and creates the ionosphere through direct ionization. I-T models use proxies for this radiation, such as Sunspot Number or the F$_{10.7}$ index.  These proxies are used in part because they are long-running history and continuity.  However, observations during the recent solar minimum have suggested that their utility may not extend to periods of extremely low solar activity \cite{Emmert:2010, Klenzing:2011, Solomon:2013}.  Additionally, their variability over the 27-day solar rotation cycle shows significant deviation \cite{Chamberlin:2007hj}.

This project should be assessed at AUL 1, as it uses existing published scientific knowledge to present a new idea for improving a specific group of space weather applications.  To advance to an AUL of 2 or higher, the project developers need to work with I-T model developers to determine if an improved solar EUV forcing index is viable and feasible for improving specific applications (for example, satellite drag calculations) and specify the level of improvement required for the application. Multiple applications could be identified from this AUL 1 project, Figure \ref{fig_branch}, each with different requirements.

\subsection{An application for another researcher: Phase II AUL 5 project}\label{Ryan}
\noindent{\bf R. M. McGranaghan } 

The following is a summary of a phase 2 AUL 5 example from the ionospheric community. This example shows a modeling project which has another research team as a user. A full version of this example can be found in Appendix \ref{append:Ryan} 

In this example the user is a researcher using the Assimilative Mapping of Ionospheric Electrodynamics (AMIE) \cite{Richmond_1988} which requires an ionospheric conductance model to infer global polar maps of electrodynamic variables on a roughly 1.5$^\circ\times$10$^{\circ}$ latitude$\times$longitude grid at variable time resolution, where the time resolution is dependent on the cadence of input observations. As researchers in the ionospheric and magnetospheric communities need event specific outputs of electrodynamic fields run on demand, the operating environment is the researcher's local computer and the relevant context changes for the specific research application. The metrics needed for this application have been outlined in \citet{Cousins:2014} and \citet{McGranaghan_2016}, in which the accuracy of the conductance model is determined by the extent to which it provides consistency between AMIE output using two different sets of input observations. 
 
All milestones through to, and including those for an AUL 5 assessment have been already completed. However, the potential performance improvement to the AMIE model, provided by the conductance model, has not been completed, and so the requirements for an AUL 6 assessment have not been met. Therefore, this project should be assessed at AUL 5 for the identified application.

\subsection{Branching applications identified through continued communication with users during development: Phase II AUL 5 project}\label{Adam}
\noindent {\bf  A. C. Kellerman}

The following is a summary of a phase 2 AUL 5 example from the magnetospheric community. This example shows a modeling project which has an industry user. It also serves to demonstrate that while a project may be at a higher AUL for one application, even a small change in the needs of the end user necessitates that the AUL be reassessed, and likely reverted back to an earlier level (see Figure \ref{fig_AUL_spiral}). It is then necessary to work through each level again to ensure that the new needs can, and will be met in a systematic and robust manner. A full version of this example can be found in Appendix \ref{append:Adam}. 

A fast real-time data assimilative framework has been developed, to produce a hindcast/forecast of the Earth's radiation belt electron dynamics. In 2016, contact was made with an external business that was interested in utilizing the real-time hindcast data to provide a real-time tool for determining if recent spacecraft anomalies were a result of space weather. A personal meeting was set up and the needs/requirements were discussed with the interested party. In this initial interaction, the need was to provide a regular output of the hindcast and forecast electron PSD and to provide a full explanation of the content and format of the available files. Through these interactions, all milestones through AUL 5 should be considered satisfied. However, since complete validation has not been completed and disseminated to the community, this project should not be assessed at AUL 6.

After further discussions with the user, there was a new requirement that errors be included with the hindcast and forecast. Since this information was not readily available at the time, the project should be assessed at AUL 2, as the same user and communication channels exist, the project has a new formalized application, and there are new requirements that address the needs of the user. Note that a revision of the users' needs necessitates the definition of a new application, as the metrics/requirements of the project have changed, and one may no longer compare the project AUL for this new application with the previous one. 

\subsection{Transitioning science to a government user: Phase II AUL 5 project}\label{Brett}
\noindent{\bf B.A. Carter and M. Terkildsen}

The following is a summary of a phase 2 AUL 5 example from the ionospheric community. This example shows a modeling project which a government user. This example also shows how one can work towards completing higher milestones while working to overcome others at lower levels. A full version of this example can be found in Appendix \ref{append:Brett}. 

A series of recent and ongoing studies into modeling the occurrence of EPBs using the Thermosphere Ionosphere Electrodynamics General Circulation Model (TIEGCM) -- a global coupled 4-D model of the ionosphere-thermosphere system \cite[and references therein]{Qia14} -- are discussed here in the context of the AULs where the user is Australia's Bureau of Meteorology (BoM). The Bureau of Meteorology's Space Weather Services (BoM-SWS) is Australia's sole provider of space weather products and forecasts. Therefore, an ongoing collaboration with BoM-SWS has helped bridge the communications and knowledge gaps between researchers and potential users of scintillation forecasts. This project should be assessed at AUL 5 as all required milestones for this level have been met. 

In the research conducted, the Raleigh-Taylor (R-T) growth rate threshold between EPB days and non-EPB days was chosen by eye to be $0.4\times 10^{-3}s^{-1}$ for several geographically separated stations, which may not be appropriate. One further complication is that the R-T growth rate is not a measurable quantity, so it cannot be directly verified against direct observations. As such, a more rigorous and systematic analysis that investigates the TIEGCM R-T growth rate threshold is needed before the milestones of AUL 6 are achieved. 

Initial work is being done towards milestones in phase III of the AUL framework. The scintillation forecasting scheme used by \citet{Car14c} is being translated into an operational environment, with the intention of providing `beta' scintillation forecasts for key users. Proceeding with the development of a working prototype and delivering forecasts has the added benefit of informing/educating the users of potential vulnerabilities to their system(s). However, while one can work toward milestones at higher levels, only once those at lower levels are met can the project be assessed at those higher AULs.

\subsection{Validating in an operational environment for multiple users, industry and government: Phase II AUL 6 project \label{Tim}}
\noindent{\bf  T. Guild}

The following is a summary of a phase 2 AUL 6 example from the magnetospheric community. This example shows a modeling project an industry and government user. It also provides an example of how a state of the art and "operational" tool may not be at an AUL 9.  A full version of this example can be found in Appendix \ref{append:Tim}. 

The SEAES tool grew out of a need to quickly assess the likelihood of the space environment causing a satellite anomaly. It has been developed at The Aerospace Corporation \cite{Koons1988, Obrien2009}. The SEAES algorithms produce a hazard quotient. This environment/anomaly likelihood relationship is derived from associating historical anomalies or their proxies to space environment measurements on the same satellites. The key user requirement for SEAES is speed:  providing a hazard quotient, or likelihood that an anomaly on their satellite is due to space weather, in near-real-time to influence decisions made.

SEAES completed tasks that satisfy Phase 1 milestones by working closely with the satellite operators during anomaly investigations where the space environment's role needed to be determined.  The interaction with users informed the development of a prototype application, outlined the requirements, and culminated in a published description of the algorithms in \citet{Obrien2009}.  The SEAES algorithms have been implemented in an operational environment at NOAA/SWPC and are available to SWPC users at the following link:  https://www.swpc.noaa.gov/products/seaesrt. These activities meet the requires of all milestones up to and including AUL 6. However, the SEAES application has never been validated in the user environment, the results have been disseminated to the users and relevant community. Therefore, this project should be assessed at AUL 6.  

\subsection{Transformative and translational research identified by the needs of the user: Phase III AUL 7 project}\label{Carl}
\noindent {\bf C. J. Henney} 

The following is a summary of a phase 3 AUL 7 example from the solar community. This example shows a modeling project for a government user. This example also shows how two projects when combined for a new application initially start at an AUL1 and must advance through all AULs with the new requirements.  A full version of this example can be found in Appendix \ref{append:Carl}. 

The ADAPT (Air Force Data Assimilative Photospheric Flux Transport) project \cite{Arge10, Arge11, Hick15} provides a sequence of best estimates of the instantaneous global spatial distribution of the solar photospheric magnetic field as a function of time. Initiated in 2008 and driven by community user interests, the objective of the ADAPT project combined two projects \cite{Word00, Hick15}, to produce global magnetic maps with realistic estimates of the uncertainty ( milestones a-c of AUL 1). An essential element during development of ADAPT has been the vital feedback and collaboration with active users ( milestones a-c in AUL 2) to assess the viability of the global maps ( milestones a - d AUL 3). Another set of fundamental steps was to integrate the ADAPT software within a prototyping environment ( milestones a-b in AUL 4), iterate on map quality, and meta-data improvements (milestones a-b AUL 5). The ADAPT model has been running at the National Solar Observatory for 5 years, generating public global magnetic maps for user validation ( milestones a-c AUL 6). The core functionality is installed at NOAA/SWPC (early stages of AUL 8 a-b) to be validated for driving WSA-Enlil, in collaboration with the CCMC. Therefore, this project should be assessed at AUL 7.

\subsection{Identifying new applications and research projects from previous targeted research: Phase III AUL 9 project}\label{Cid}
\noindent {\bf C. Cid} 

The following is a summary of a phase 3 AUL 9 example from the ground induced current community. This example shows a data project for a government user. This example also shows how through completion and continued validation of a project assessed at AUL 9 can lead to the funding of new targeted basic research projects. 

After the October 2003 Halloween Storm, affecting electric utilities in South Africa \cite{Kappenman2005}, low, and mid geomagnetic latitude countries were made aware that their power grids might be vulnerable to this hazard. The magnitude of the Spanish geomagnetic latitude is similar to South Africa. This fact was the impetus for a chain of projects that concluded with the development of a new index for nowcasting geomagnetic disturbances and the geomagnetically induced currents risk in Spain.

This project has met the requires of all the milestones and to and including those of AUL 9, and should be assessed at that level. LDi and LCi products have been implemented through the ESA SSA Space Weather Service Network (http://swe.ssa.esa.int). The validation of the products continues while working in an operational environment. However, this is not the end of the story, new research goals were formed to answer the questions opened during the previous project. Now that we know that the new geomagnetic indices LDi and LCi are useful for nowcasting the disturbances at the ground level for local users, our aim is to forecast these indices from solar wind data. Revising the performance of the models that forecast ground magnetic disturbances from solar wind data, we discovered that they provide good results for smooth changes in local ground records, but not for fast changes, which are the most relevant for power grid users. To achieve this new goal, we need to understand the complex physics that solar wind-magnetosphere interactions rely on during transient phenomena. Some steps have already been taken \cite{Saiz2016} and a basic research project, funded by the Spanish government, is ongoing and now, and should be assessed at AUL 1.

\subsection{Funding an application's progress through the three phases: Phase III AUL 9 Project \label{NYG}}
\noindent {\bf N.Yu. Ganushkina} 

The following is a summary of a phase 3 AUL 9 example from the magnetospheric community. This example shows a modeling project for researcher and government users. This example also shows how projects may be funded as they go through different phases.  A full version of this example can be found in Appendix \ref{append:NYG}. 

The Inner Magnetosphere Particle Transport and Acceleration Model (IMPTAM) was developed for purely scientific purposes (Ganushkina et al., 2000, 2001). IMPTAM moved towards applications when the project called SPACECAST was funded in 2011. The main goal of SPACECAST was to protect space assets from high energy particles. The identified users for IMPTAM were the BAS (British Antarctic Survey, Cambridge, UK) and the ONERA (Office National D'Etudes et de Recherches Aerospatiales, Toulouse, France). The first nowcast version of IMPTAM for $<$ 100 keV electrons (Ganushkina et al., 2013, 2014) running online in real time (http://fp7spacecast.eu/) was developed. Validation of the IMPTAM output has been ongoing since the initial operation online (Ganushkina et al., 2015). The project continued in collaboration with SPACESTORM (http://www.spacestorm.eu/) and at the completion of activities that satisfy the milestones up to and including AUL 6. The IMPTAM fully was integrated and implemented at the user's system, for the given application. All milestones up to and including those for AUL9  are met, and this project should be considered AUL 9 for this application. 

At present, IMPTAM is in another AUL spiral, Figure \ref{fig_AUL_spiral}, as part of the on-going project PROGRESS funded by the European Union's Horizon 2020 research and innovation programme. In this project, IMPTAM is undergoing transformations to operate as a predictive tool.



\section{Summary and Discussion}
\label{summary}
The proliferation and variety of models and data sets is a sign of the advancement and complexity of our field. As our field becomes more inter, multi, and transdisciplinary, researchers need to provide a wide array of information, data, and predictions. Our field now regularly encompasses collaborations including coupling and interactions across the heliosphere, data collection for Earth and astrophysical sciences, and applications to planetary and exoplanetary environments. Communication within our field and with related research areas is crucial to the effectiveness of our research outcomes. Similarly, there is a strong need for better communication with communities outside of academic research. One necessary step for R2R, R2O, and O2R  is recognizing the needs of the user and how those needs inform requirements. This includes identifying the most useful metrics, accuracy, and format of information. As technology becomes more susceptible space weather, the challenge for researchers is to communicate clearly to users the capabilities of their research and how it can be beneficial for their needs \cite[e.g.,][ and references therein]{Baker2008a, Caldwell2017, Cassak2017}. 

To further enable communication of a project's progress towards defined outcomes, we have proposed the Application Usability Level framework. The AUL framework provides a step-by-step approach for tracking a project's progress towards a specific application. The framework we have presented here is intended for communication both within basic research fields and with industry users. As such, we have framed this work around two populations: {\it researchers} and {\it users}.  The users may refer either to non-academic users or to other researchers. However, in the case of non-academic fields, users and researchers alike may benefit from a translator, i.e. a {\it broker}, who may help with the effective transition of research to operations or from one research field to another. In such cases, the expected users may not have the means or resources to fully explore the possibilities of a given model or data product for their application. Likewise, researchers might not be best positioned to appreciate the user's needs.  Independent subject matter experts can be critical as brokers. They can ensure that the AULs are developed and tailored correctly for a given project. Brokers can include forecasters at government agencies (e.g., NOAA and the UK Met Office), government and government-funded scientists (including FFRDCs and government labs), academics or industry partners. Brokers should be sought out as needed early in the AUL process. In many cases, the brokers will become the user for many AUL pathways.

Within the AUL framework, the validation needs and subsequent definition of metrics are set early in the process (AUL 2). While the framework described in this paper applies to individual users with specific needs, the space physics community as a whole has a role to play in enabling the discovery and viability testing in Phase 1. The definition of a standard set of metrics for a given application, such as the CCMC's CME arrival scoreboard, can simplify the process in Phase 1. This can help ensure that each user is applying the right tool for the job. The validation and metric needs in the case of benchmarking involve the uniform application of metrics across different data or model frameworks which can measure improvement over time. The Community Coordinated Modeling Center (CCMC) has a unique role in helping to define and retain standards for use across the Heliophysics community.  These efforts can be found through the work done in the International Forum for Space Weather Capabilities Assessment working groups (For more information on, and to get involved with these efforts see https://ccmc.gsfc.nasa.gov/assessment/forum-topics.php and other papers in this special issue). Community efforts such as Coupling, Energetics, and Dynamics of Atmospheric Regions (CEDAR),  GEM, and Solar Heliospheric and INterplanetary Environment (SHINE) also play a role in testing these metrics. They provide an arena to test their ease of application, their usefulness for a constantly evolving scientific community, and in employing the standard metrics with cutting-edge models that may not yet be available at the CCMC.

As implied above, previous large scale community-driven efforts have focused on the validation of models and cross-calibration of instruments. These community efforts have been vital to the progression of our field but have often been centered around the needs of the researcher. The initial vision for the GEM workshop was to create multiple magnetospheric modules that would eventually be combined to produce a comprehensive model of the geospace environment and its interactions with the solar environment \cite{Roederer1988}. Efforts like the GEM Workshop \cite[e.g.,][]{Raeder1998, Birn2001, Jordanova2006, Rastatter2013}, the CCMC \cite[e.g., ][]{Bellaire2006, NAP10477,NAP13060}, and the Center for Integrated Space weather Modeling (CISM) \cite[e.g., ][]{SPENCE2004, NAP13060} were instituted to enable coordination and intercommunication within and between codes. Specifically, these groups encourage the coupling of codes solving for different regions of space for the purpose of predicting the properties and variability of the space environment. The CCMC has played a crucial role in making these models available to the public. The CCMC has further helped in the validation of models and communication with users outside the space physics community. These and many other efforts continuing to make strides in improving and validating predictive models, defining metrics, and enabling communication within the field \cite[e.g.,][]{Owens2008, Quinn2009, Shim2012, Honkonen2013, Pulkkinen2013, Rastatter2013, Gordeev2015, Glocer2016}. The AUL framework can help with identifying and providing the data products for inputs into these models as shown in example \ref{Jeff}. It can also help with the coordination of coupling models, as shown in examples \ref{Ryan}, and \ref{Carl}.  And finally, the AUL framework can inform industry users and forecasters as to the usability of the project as demonstrated in examples \ref{Brett}$-$\ref{Cid}, and \ref{NYG}. 

In this paper, we outlined the AUL framework and defined the different phases, levels, and the milestones necessary to reach each step for a project's AUL. We discussed potential methods of disseminating results as well as best practices. Several example summaries were provided, and full examples can be found in the appendix, \ref{append}, which shows how this framework can be applied to current projects as well as how working with users can lead to scientific discoveries and future projects. Development of the AUL framework began at the first working meeting for the International Forum for Space Weather Capabilities Assessment by members of the Assessment of Understanding and Quantifying Progress working group. The aim of this working group is to develop a framework to aid in tracking the progress of our field and to provide a path for clear communication between researchers, funding agencies, and users.

\appendix
\section{Longer version of example projects using the AUL framework}
\label{append}
Within the appendix we provide longer more explicit versions of the examples within the primary text of the paper. 

\subsection{Identifying a potential new application to track with the AUL framework: Phase I AUL 1 project}\label{append:Jeff}
\noindent {\bf  J. Klenzing and A. G. Burrell}

The use of average solar extreme ultraviolet (EUV) as a potential driver of ionosphere-thermosphere (I-T) models is an example of an early-phase AUL 1 project.  A brief outline follows for the start of a project set to use the AUL framework. 

The specific EUV radiation that is used as a fundamental driver of I-T models is the spectra from 0.05$-$105.0 nm.  This radiation heats the thermosphere and creates the ionosphere through direct ionization. Historically, I-T models have used proxies for this radiation, such as Sunspot Number (SSN) or the F$_{10.7}$ index.  These proxies are used in part because of the long-running history and data continuity.  However, observations during the recent solar minimum have suggested that the utility of these proxies may not be extended to periods of extremely low solar activity \cite{Emmert:2010,Klenzing:2011,Solomon:2013}.  Additionally, the variability of these proxies over the 27-day solar rotation cycle shows significant deviation \cite{Chamberlin:2007hj}.  Because the solar atmosphere can have different transmission properties for wavelengths of nm (EUV) vs cm (F$_{10.7}$), the day-to-day variability of these parameters can match quite well at times while being substantially different at other times.  

This project is assessed at AUL 1, since it uses existing published scientific knowledge to present a new idea for improving a specific group of space weather applications.  
To advance to an AUL of 2 or higher, the project developers would need to work with I-T model developers to determine if an improved solar EUV forcing index is viable and feasible for improving specific applications (for example, satellite drag calculation and collision avoidance due to thermospheric heating) and specify the level of improvement required for the application.  Potentially, multiple AUL 2 applications could be identified from this project, as shown in Figure \ref{fig_branch}, each with different requirements.

\subsection{Example of a Phase II, AUL 5 project - R. M. McGranaghan }\label{append:Ryan}

 \subsubsection{Application: Conductance models to calculate high-latitude ionospheric electrodynamic fields}
In this example the end user is the Assimilative Mapping of Ionospheric Electrodynamics (AMIE) \cite{Richmond_1988} which requires an ionospheric conductance model to infer global polar maps of electrodynamic variables (electric and magnetic fields and horizontal and field-aligned currents) on a roughly 1.5 $x$ 10$^{\circ}$ latitude $x$ longitude grid at variable time resolution, where the time resolution is dependent on the cadence of input observations. As researchers in the ionospheric and magnetospheric communities need event specific outputs of electrodynamic fields run on demand, the operating environment is the researcher's local computer and the relevant context changes for the specific research application. 

Metrics to validate the conductance model are challenging due to the fact that conductance cannot be directly measured. However, the metric needed for this application has been outlined in \citet{Cousins:2014} and \citet{McGranaghan_2016}, in which the accuracy of the conductance model is determined by the extent to which it provides consistency between AMIE output using two different sets of input observations (i.e., space-based magnetic perturbation observations from AMPERE and ground-based ionospheric convection observations from SuperDARN). The specific bounds in the error of the AMIE procedure due to the conductance model will depend on the goals of the specific research application. There are a number of efforts currently working on providing the necessary inputs to the AMIE model which are described below.
 
\subsubsection{GLobal airglOW (GLOW) model Soloman et al 1988 }
The conductance model makes use of the GLOW electron transport and upper atmospheric chemistry model with specification of the auroral particle precipitation by Defense Meteorological Satellite Program (DMSP) satellites observations to calculate in-track conductance estimates, and then performs an assimilation of these in-track data to obtain global high-latitude conductance distributions \cite{McGranaghan:2015, McGranaghan_2016}.
 
For this application, one set of users are research modelers using the AMIE procedure introduced above. The GLOW model is currently being validated for various geomagnetic conditions which cover the environmental conditions necessary for this specific application. However, the assimilative GLOW model has already been examined for a characteristic event containing several periods of both quiet and geomagnetic storm and shown to provide greater accuracy than conductance models currently in wide use \cite{McGranaghan_2016}. The GLOW model will be able to provide the conductances necessary to run AMIE in the required environments for the end user and thus this project should be assessed at AUL 1.
 
This application has been deemed feasible for the research application by both the end user researchers as well as the developers. The assimilative conductance model is capable of providing the necessary conductances at the spatial and temporal resolutions required by the AMIE model and can be run on demand in a timely manner. The model has been tested and validated \cite{McGranaghan_2016} and the detailed characterization of the baseline performance and limitations have been completed. Thus, all milestones in the Phase I development are met, and so this project can be assessed at AUL 3.
 
Phase II of the AUL framework is focused on the development, testing, and validation of the conductance model for the application of providing on-demand conductances for input into the AMIE model for research purposes. The model is running in the operational environment and has been demonstrated to be able to be run on demand for the end user needs. The organizational challenges have been managed. These activities satisfy the AUL level 4 milestones, and this project can be assessed at AUL 4.

Future efforts to validate the global conductance patterns from the assimilative GLOW model will involve systematic testing across different relevant contexts. Additionally, ongoing efforts are comparing GLOW model output to other conductance models in various forms \cite[e.g.,][]{Grubbs:2018} and may lead to new metrics for this application.

The potential improvement upon the state of the art has been articulated \cite{McGranaghan_2016} and the capability to run the model during the relevant context conditions necessary for the research (quiet and storm time) has been developed. During both quiet and storm conditions the model can meet the requirements needed for the AMIE procedure, and, thus, milestones and Levels through AUL 5 have been fulfilled. Finally, because current efforts are underway to demonstrate the potential performance improvement to the AMIE model provided by the conductance model, requirements for AUL 6 have not yet been completed. Therefore, this project can be assessed at AUL 5 for this application.

\subsection{Example of a Phase II project: AUL 5 - A. C. Kellerman}\label{append:Adam}

\subsubsection{Hindcasting and Forecast Radiation Belt Electron Fluxes}

The Versatile Electron Radiation Belt (VERB) code \cite{Sub09} has been recently combined with a Kalman filter \cite{Kal60}, data from the Van Allen Probe MagEIS \cite{Bla13} and REPT \cite{Bak13b} instruments, and data from the GOES MAGED and EPEAD instruments in order to develop a data-assimilative code \cite{Shp13c,Kel14}. The computational requirements for a full three dimensional Kalman filter may be quite large in the domain required for radiation belt simulations an alternative \textit{split-operator} approach was introduced \cite{Shp13c}. The data-assimilative code was applied to study the March 1991 superstorm, leading to the discovery of a 4-zone structure in the Earth's radiation belts, identification of local acceleration events during a historical geomagnetic superstorm, and the development of the first data-assimilative radiation belt forecast model \cite{Kel14}. The forecast model runs at UCLA, has been in operation since February 2015, and has recently been adapted to provide output for users, outside of the research community.

The on-line forecast model is largely a research model, adapted to run automatically every two hours, producing a hindcast and a forecast. The hindcast assimilates available spacecraft observations, in this case, real-time GOES primary and secondary data, and the real-time Van Allen Probes MagEIS and REPT data. The forecast utilizes the VERB code and forecast Kp to predict the change in electron phase space density (PSD) across multiple values of the three adiabatic invariants \cite{Roe70,Sch74}. 

\subsubsection{Phase I}

The first step required to take the project forward was to identify how this tool may be used for decision making or a particular application. The forecast model provides a recent hindcast of the state of the Earth's radiation belts and a forecast of the state up to two days in the future. Several recorded failures of spacecraft electrical systems have been reported in the past as a result of geomagnetic activity. One such example is the failure of the attitude control system on the Galaxy 4 spacecraft in 1998 \cite{Bak98}. In order to determine whether a recent failure is due to geomagnetic activity, it may be useful to have a real-time monitor of the radiation environment, which provides information to operators. The identification of this application {\it satisfies milestone a) in Phase 1, AUL 1}.

The development of the VERB code 2 has been documented in the literature \cite{Sub09}, the code has been tested for numerical accuracy \cite{Sub09}, and validated against spacecraft observations \cite{Sub11b,Kim12}. The data assimilative model was tested for sensitivity to the datasets included in \citet{Kel14}. Current documentation of the code is available to all users, and there is an extensive set of examples which provide ease of access to the code. Both of these products are available on request and are maintained on a dedicated Gitlab server. The data assimilative aspect of the code has been tested and published in the literature \cite{Shp13c,Kel14}, and example scripts to load the reanalysis and conduct investigations are available also via a Gitlab server. {\it Together these items satisfy milestone b) in Phase 1, AUL 1}. 

Over the past few years, contact has been made between researchers and users who may be interested in the real-time operational forecast model output. Most of the contact occurred at national and international conferences, and through email exchange. {\it This list of potential users addresses milestone c) in Phase 1, AUL 1, and hence the project should be assessed at AUL 1}

Contact was made with business who was interested in utilizing the real-time data to determine potential risks to spacecraft as a result of deep dielectric charging \citet{Meu76}. A personal meeting was set up and the needs/requirements were discussed with the interested party. In this initial interaction, the need was to provide regular output of the hindcast and forecast electron PSD and to describe the file format and any other information associated with the data contained in the files. Email and telephone methods for communication were set up, and a schedule of activities/contact times were agreed upon. {\it These conditions address milestones a) and b), Phase 1, AUL 2}

At this stage, the application has been identified - An industry user who wants to develop risk-assessment software, and at this stage just wants to begin using the data for testing.  It was also decided at this stage that it would not be feasible (or necessary) to implement the model remotely, and so the implementation would remain at UCLA. Therefore, the only metric for success was to ensure a sufficient storage device, 24/7 access, and some file format changes for brevity. \textit{With the metrics for success identified, milestone c), Phase 1, AUL 2 is complete, and the project should be assessed at AUL 2}.

At the initiation of the project, no other data-assimilative radiation belt forecast model was in existence, and hence the project can be considered state of the art. In the given application, the research model may be used to inform decision making for the satellite industry, and there will be a user who will regularly look at the output from the model. Both of these are beneficial for the research direction, as feedback will be received that may help to inform model development in the future. As mentioned previously, there were only a few modifications necessary to achieve success, in terms of AUL ascension, for this project, which likely would provide great feedback to research and development, hence the project is considered viable and implementation will be feasible. \textit{milestones a)-c), AUL 3, Phase 1 is complete}. The project should be assessed at AUL 3.

\subsubsection{Phase II}

The user requires a model that can provide information for determining whether deep dielectric charging may have occurred for a recent anomaly. The forecast version of the data-assimilative VERB code was developed largely from the published version \cite{Kel14}, and implements the validated and tested VERB code \cite{Sub09,Kim12,Dro15,Ase16} to provide model-matrices for the assimilation framework. The code was implemented into an operational framework, capable of producing electron PSD nowcasts and forecasts. All organizational challenges were overcome, and the project has been {\it integrated into the environment required by the application, satisfying milestones a) and b), AUL 4, Phase 2}. The project should be assessed at AUL 4. 

The model was set up to run automatically, producing documented output files required by the user every 2 hours. These files are currently being loaded remotely by the user on a regular basis. The application has been integrated into a functioning application system, providing state-of-the-art estimates of electron PSD in real time. The project should be assessed at AUL 5.

The application has been tested first in a simulated environment, and now in an operational environment. Documentation and further testing is still underway for this project, and hence it can not be rated at AUL6.

\subsubsection{Back to Phase I}

Since implementing the project, there have been further discussions with the user. There is a new requirement that errors be included with the hindcast and forecast electron PSD. A data product with known errors requires further investigation and development of the model. For this new application, the project should be rated at AUL 2, as the milestones for AUL 3 require some further work before they can be considered complete. Note that a revision of the users needs necessitates that one treats the application as a new application, as the metrics/requirements have changed, and one can no longer compare the project AUL for this new application with the older one.

This example serves to demonstrate that while a project may be at AUL 5 for one application, a small change in the needs of the end user necessitates that the AUL be reassessed, and likely reverted back to an earlier level. It is then necessary to work through the levels again to ensure that the new needs can, and will be met in a systematic and robust manner. 


\subsection{Example of a Phase II project: AUL 5 - B.A. Carter and M. Terkildsen}\label{append:Brett}
\subsubsection{Ionospheric scintillation prediction}

Equatorial Plasma Bubbles (EPBs) are low plasma density structures that rise up into the high plasma density in the Earth's ionosphere during the nighttime hours \citep[e.g.,][]{Kel11}. EPBs, also known as Convective Ionospheric Storms, generate a spectrum of plasma waves/irregularities that cause random fluctuations (i.e., ``scintillations'') in the amplitude and phase of radio waves that propagate through them; e.g., those used for Positioning, Navigation and Timing (PNT). The amplitude and phase scintillation can cause Global Navigation Satellite System (GNSS) receivers to loose lock with one or more satellites, which can adversely impact the PNT results. The impact of ionospheric disturbances was highlighted in the 2015 multi-agency Space Weather Action Plan (National Science \& Technology Council, 2015 \cite[and references within]{SWAP}, and NASA's Heliophysics Living with a Star Program identified ``Physics-based Scintillation Forecasting Capability'' as one of seven Strategic Science Areas in their recent decadal plan \citep{NASA2015}. Therefore, a current focus of the ionospheric research community has been to understand the driving mechanisms of the growth of EPBs, with the end-goal of developing an accurate EPB forecasting capability.

EPBs are known to be caused by the Generalized Rayleigh-Taylor (R-T) plasma instability \cite[e.g.,][]{Sul96}, in which a sharp vertical gradient in the plasma density in the bottom of the F layer, coupled with an upward plasma drift, creates instability in the ionospheric plasma. The plasma perturbations generated in the bottom-side of the F layer (approx. 150-200 km altitude) undergo rapid nonlinear growth into large `bubbles' of low-density plasma that rise towards the topside of the F layer.

Empirical and phase screen propagation models have been shown to be very useful in not only capturing the EPB occurrence climatology but also reproducing typical scintillation levels that are observed on the ground; the Wideband ionospheric scintillation model (WBMOD) \citep{Sec95} and the Global Ionospheric Scintillation Propagation Model (GISM) \citep{Ben11} are notable examples. On the other hand, the development of a physics-based prediction capability for EPBs has two primary challenges. Firstly, physics-based predictions must have the ability to predict the occurrence of EPBs by simulating the background ionospheric conditions; particularly the daily changes in the upward plasma drift near sunset. The next challenge is to then numerically model the formation and small-scale structure(s) of the EPBs themselves \cite[see review by][]{Yok17}. In achieving both of these challenges, an EPB prediction capability would be able to forecast the occurrence of EPBs, their spatial extent and their impact on radio waves. The example discussed here deals with the first of these challenges; i.e., daily EPB occurrence.

A series of recent and ongoing studies into modeling the occurrence of EPBs using the Thermosphere Ionosphere Electrodynamics General Circulation Model (TIEGCM) -- a global coupled 4-D model of the ionosphere-thermosphere system \cite[and references therein]{Qia14} -- are discussed here in the context of the AULs.

\subsubsection{Phase I}

AUL 1: Basic Research

While GNSS and Satellite Communications users are now widely recognized as the primary end-users, EPB prediction has been a topic of significant research effort since before the Global Positioning System was deployed. As such, the community benefits from decades of basic research into what began as a pure scientific curiosity into ``Equatorial Spread F'' \citep{Boo38}. Basic research into EPBs is an ongoing topic, but the physical mechanism that drives the generation of EPBs is well-understood to be the R-T plasma instability.

The field has been working with the rationale that scintillation event forecasts would be useful for both Satellite Communications and GNSS users. This AUL framework would help enable communication with potential end users and help publicly track the progress of this project towards meeting their needs/requirements. Therefore, the AUL 1 milestones have been satisfied. 

AUL 2: Establishment of users and their requirements for a specific application

Currently, the global community of GNSS users spans across many key industries, and one of those is aviation, with which weather forecasting agencies, such as Australia's Bureau of Meteorology (BoM), have existing communications channels. Further, the Bureau of Meteorology's Space Weather Services (BoM-SWS) is Australia's sole provider of space weather products and forecasts. Therefore, an ongoing collaboration with BoM-SWS has helped bridge the communications and knowledge gaps between researchers and potential end users of scintillation forecasts.

From this collaboration, we've learned that both GNSS-based positioning and surveillance and the use of Satellite Communications in the aviation sector are growing rapidly in line with the move to Performance-Based Navigation. Amplitude scintillation can have a significant impact on aircraft using GNSS for Required Navigation Performance-based flight navigation. Further, ionospheric scintillation may disrupt satellite communications-based technology used by aircraft, with the potential to impact both communications and surveillance. The aviation sector, through the International Civil Aviation Organization (ICAO), has identified ionospheric scintillation (both current and forecast conditions) as one of a number of space weather information requirements for aviation users. These requirements will be formalized in ICAO Standards and Recommended Practices, as a recommendation for ICAO-nominated space weather centres to provide ionospheric scintillation advisories and forecasts to aviation users at not more than 6 hourly intervals. With these requirements in mind, the milestones for AUL 2 have been satisfied.

AUL 3: Assess viability of concept and current state of the art

Currently, scintillation products are built around recent or current observations from ground-based GNSS receivers, or based on climatological models \cite[e.g.,][]{Sec95,Ben11}. Very few scintillation forecast products are currently openly available, and to the authors' knowledge, those that do exist are largely built around climatology or extrapolating recent conditions. Consequently, they tend to capture the seasonal climatology, but not necessarily the day-to-day variability in ionospheric scintillation, which is necessary for end users to implement effective mitigation strategies.

Therefore, an EPB/scintillation prediction that is capable of capturing daily variability could significantly advance the state-of-the-art scintillation modeling/forecasting capabilities and is the focus of the current project. The initial results of this analysis into the viability of using physics-based modeling for this purpose were published in \citet{Car14} (discussed in further detail below), thus the AUL 3 milestones have been satisfied.

\subsubsection{Phase II}

AUL 4: Initial integration and verification

\citet{Car14} was the first to use the TIEGCM to directly calculate the flux-tube integrated R-T linear growth rate derived by \citet{Sul96}. In \citet{Car14}'s work, a daily variation in the maximum R-T growth rate was revealed, and this variation showed a clear resemblance to the occurrence of amplitude scintillation, as measured using a ground-based GPS receiver. Further analysis in that study, and in a subsequent study \cite{Car14b}, examined the source of the TIEGCM R-T growth rate daily variability, and found it to be caused by variations in the TIEGCM's magnetospheric input in the high-latitude region; i.e., the electric potential patterns that drive horizontal plasma drift. The high-latitude plasma flow variations were found to influence the thermospheric winds in the equatorial region hours later, and these changes were found to influence the strength of the R-T growth rate. These thermospheric wind variations that were modeled by the TIEGCM is understood to be the ``disturbance dynamo'' effect \citep{Bla80}.

The analyses discussed above focused on the peak EPB season when EPB occurrence is dictated by conditions that suppress, not enhance, EPB growth. As such, the TIEGCM's ability to show a decreased R-T growth rate on one day compared to the day prior represents an ability to model (and potentially forecast) daily variations in EPB activity. This analysis satisfied the AUL 4 milestones.

AUL 5: Demonstration in the relevant context

To demonstrate the feasibility of employing the TIEGCM R-T growth rate results in an operational EPB prediction environment, \citet{Car14c} used the \citet{Win05} forecast Kp index to drive the TIEGCM in a 5-month EPB prediction trial for six locations across Africa and Asia. In this analysis, a threshold R-T growth rate of 0.4$\times 10^{-3}$s$^{-1}$ was used to classify whether the day would be an EPB day or a non-EPB day. During peak EPB season, it was shown that the TIEGCM R-T growth rate predictions were successful in capturing non-EPB days, as measured by the ground-based GPS receivers. 

\begin{figure}
\centering
\noindent\includegraphics[width=0.7\textwidth]{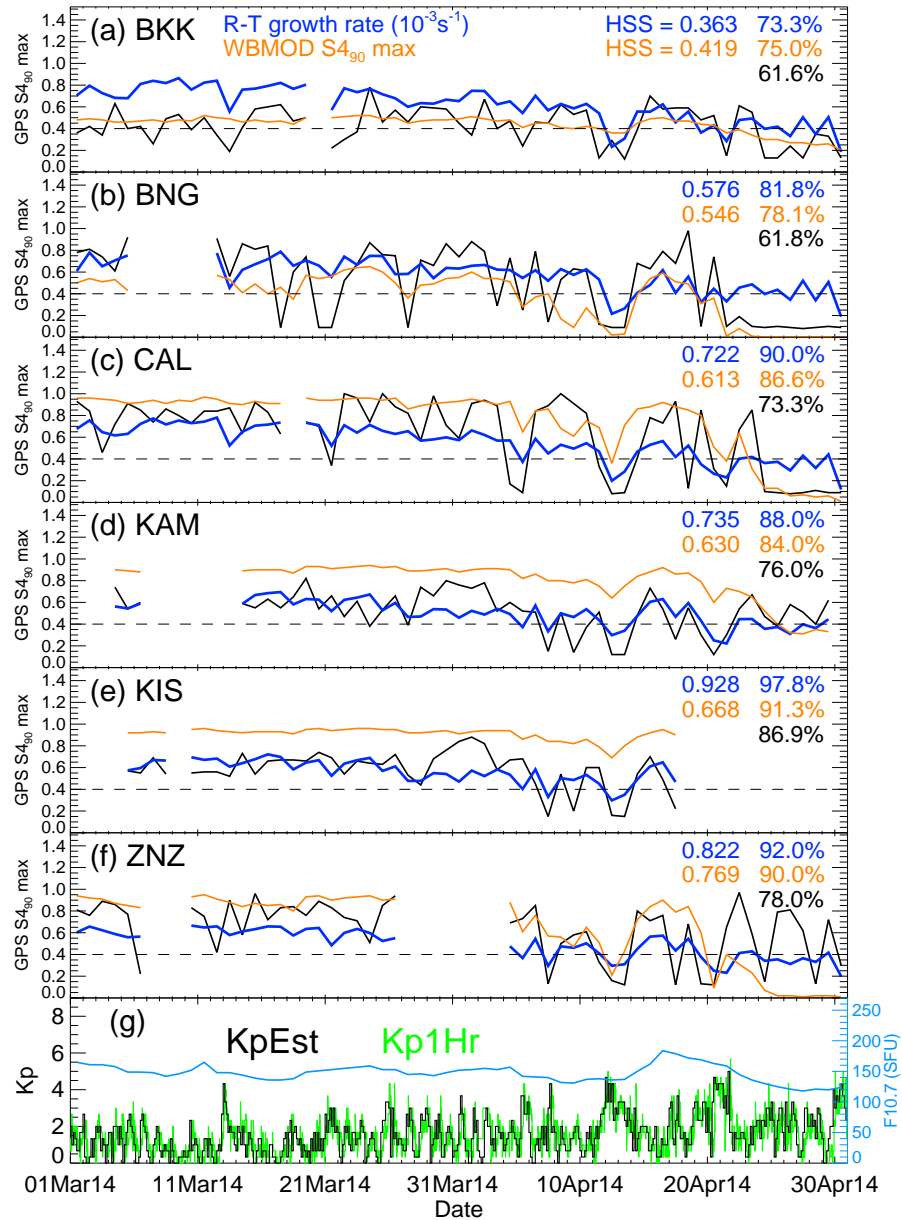}
\caption{(a) to (f) The daily GPS $S4_{90}max$ observed by each GPS station throughout March and April of 2014 in black. The orange lines show the WBMOD predictions for GPS $S4_{90}max$ and the blue lines show the TIEGCM R-T growth rate. In the top-right of each panel is the corresponding Heidke Skill Score and the percentage of correct EPB/non-EPB days forecast. The black percentage indicates the ``persistence'' forecast result. The dashed horizontal line indicates the S4 and R-T growth rate thresholds. (g) The real-time observed Kp (KpEst, black), the 1-hour predicted Kp (Kp1Hr, green) and the F$_{10.7}$ solar flux (blue) throughout this period.}
\label{Fig:EPB_SCINDA_1hr}
\end{figure}

Figure \ref{Fig:EPB_SCINDA_1hr} shows a 2-month subset of the results from the 5-month period analyzed by \citet{Car14c}. In this analysis, the amplitude scintillation S4 index, which is measured each minute for each satellite-to-ground link, is used. Each hour, the 90th percentile of the S4 index from all satellite links 30$^\circ$ above the horizon, $GPS S4_{90}$, is taken to indicate the presence of elevated scintillation activity. The black solid curves in Figures \ref{Fig:EPB_SCINDA_1hr}a-f are the daily maxima of the GPS S$4_{90}$ (GPS S$4_{90}$max) throughout March and April 2014 for three GPS station locations in Southeast Asia (Bangkok, Bangdun and Calcutta) and three locations in Africa (Kampala, Kisumu, and Zanzibar). The blue curves indicate the daily maximum R-T growth rates that were calculated from the TIEGCM. The orange curves show the predicted GPS S$4_{90}$max from WBMOD \citep{Sec95}, for comparison with state-of-the-art. The horizontal dashed lines indicate the chosen GPS S$4_{90}$max=0.4 threshold (i.e., the difference between an EPB day and a non-EPB day) and the chosen R-T growth rate threshold of $0.4\times 10^{-3}s^{-1}$. Based on these thresholds, the Heidke Skill Scores were calculated and are shown in the top-right of each panel, alongside the percentage of days with an accurate EPB occurrence forecast for each model. The black percentage indicates the success of the ``persistence'' forecast; i.e., what happened yesterday will happen today. Finally, Figure \ref{Fig:EPB_SCINDA_1hr}g shows the nowcast Kp index (KpEst, black), the 1-hr Wing Kp forecast (Kp1Hr, green) \cite{Win05} and the F$_{10.7}$ solar flux (blue).

In this demonstration, it can be seen from both assessment metrics that the TIEGCM R-T growth rate is generally better at capturing the EPB daily variability than both the ARFL WBMOD and the persistence forecasts. The non-EPB days, e.g., April 13th for all stations, are characterized with R-T growth rates of less than the $0.4\times 10^{-3}s^{-1}$ threshold. It should be noted that the TIEGCM does not capture all non-EPB days; e.g., April 9th for both KAM and KIS stations.

Importantly, as the TIEGCM was driven using the 1-hr Wing Kp index forecast, this demonstration was designed to be comparable to an operational prediction environment that would or could be used by space weather forecasting agencies. With prediction accuracies generally consistently higher than those from WBMOD, and consistently higher than the persistence forecast, the TIEGCM was shown to be useful in the prediction of EPBs on a daily basis during peak EPB season, effectively satisfying the AUL 5 milestones.

AUL 6: Complete validation

While this initial assessment is promising, there are some further questions that need answering prior to completely achieving AUL 6. In particular, the thresholds used for both the scintillation level and the R-T growth need to be further investigated. 

The primary challenge with using a single S4 threshold is that the background electron density is proportional to the scintillation level \citep{Wha09}. Therefore, stations located under the equatorial anomaly trough (i.e., at the magnetic equator) are going to register lower S4 values compared to stations located under the anomaly crests. The levels across these different locations, which are also likely to change with time (e.g., season, solar activity) need to be quantified.

In the work discussed above, the R-T growth rate threshold was chosen by eye to be $0.4\times 10^{-3}s^{-1}$ for all six stations, which may not be appropriate for all stations in different longitude sectors; \citet{Car14b}'s analysis uncovered notable differences between the optimal R-T growth rate thresholds between different longitude sectors. One further complication is that the R-T growth rate is not a measurable quantity, so it cannot be directly verified against direct observations. As such, a more rigorous and systematic analysis that investigates the TIEGCM R-T growth rate threshold is needed before the milestones of AUL 6 are achieved. 

Both of these aspects are part of ongoing work.

\subsubsection{Current work and future plans}

Firstly, current research is focused on quantifying the optimal/most reliable scintillation and R-T growth rate thresholds for given locations, and exploring the conditions under which these thresholds should be adapted. Verification of scintillation forecast products in terms of ground-based S4 estimates is straight forward and a good amount of high-quality S4 data exists for this purpose. Many space weather agencies are also using proxies for ionospheric scintillation such as ROTI, and these show good correlation with scintillation indices.

Also, some initial work is being done towards achieving AUL 7 (i.e., Application prototype), in collaboration with the Australian BoM-SWS, and in consultation with the TIEGCM developers (National Center for Atmospheric Research, NCAR). The current goal is to set up the scintillation forecasting scheme used by \citet{Car14c} in an operational environment, with the intention of providing `beta' scintillation forecasts for key end-users (such as aviation). 

While it may seem premature to proceed with tasks that would satisfy AUL-7 milestones, without having completed the validation necessary for an assessement of AUL 6, we expect that any findings related to the scintillation and modeled growth rate thresholds could be easily translated into an operational `beta' scintillation forecasting system. 

Proceeding into the user-system implementation will help with the challenge of verifying the scintillation forecasts and advisories in terms of end user experience, primarily because end user experience varies with such things as application, equipment, usage, and tolerance. Proceeding with the development of a working prototype and delivering forecasts has the added benefit of informing/educating the end users of potential vulnerabilities to their system(s). This direct interaction with end users will also hopefully create a feedback loop that will allow for modifications to an operational scintillation forecasting system in order to make it more useful/informative for them. BoM has a regular program of engagement with the aviation community providing valuable feedback on pilot products and services. These interactions will thus help to satisfy AUL-8 milestones (Validation in relevant ``real world'' environment, and eventually AUL-9 milestones (Approved for on-demand use towards stated application).

Looking further forward, it is worth mentioning that research on global physics-based ionosphere-thermosphere modeling is continuing to advance; e.g., the most recent release of WACCM-X \citep{Liu17}. Further, data assimilation is being investigated as a tool for capturing daily R-T growth variability \cite[e.g.,][]{Raj17}, and ground-to-topside modeling has been used to show that lower atmospheric forcing can be a significant source of daily variability in the R-T growth rate \citep{Shi18}. Thus, research into using global ionosphere-thermosphere models for predicting EPB occurrence is expected to continue to adapt as these models and techniques continue to be expanded upon and improved.

\subsection{Validating in an operational environment for multiple users, industry, and government: Phase II AUL 6 project \label{append:Tim}}
\noindent{\bf  T. Guild}

The SEAES tool grew out of a need to quickly assess the likelihood of the space environment causing a satellite anomaly.  It was originally developed at The Aerospace Corporation \cite{Koons1988} and modernized by \citet{Obrien2009}.  The SEAES algorithms produce a hazard quotient, which is the ratio of the instantaneous likelihood of an anomaly to its long-term mission-averaged likelihood of an anomaly.  This environment/anomaly likelihood relationship is derived from associating historical anomalies or their proxies to space environment measurements on the same satellites, yielding a translation between environment and hazard.  The key user requirement for SEAES is speed:  providing a hazard quotient, or likelihood that an anomaly is due to space weather, in near-real-time to influence decisions made during satellite anomaly investigations. 

SEAES completed activities that satisfy the AUL 1-3, Phase 1, milestones during the early development at Aerospace, working closely with satellite operators during anomaly investigations where the space environment's role needed to be determined.  The interaction with users informed the development of a prototype application, outlined the requirements, and culminated in a published description of the algorithms \citep{Obrien2009}.   This work satisfies all of the Phase 1 AUL milestones, and this project should be assessed at AUL 3.

This prototype application has been implemented in a relevant DOD computing network to facilitate delivering hazard quotients to users, and feedback to the development team.  In addition, the SEAES algorithms have been implemented in an operational environment at NOAA/SWPC and are available to SWPC users at the following link:  https://www.swpc.noaa.gov/products/seaesrt.  This satisfies Phase 2, through AUL 6 milestones.  However, the SEAES application has never been thoroughly validated within the user environments. The results been disseminated to the user and relevant communities.  We can, therefore, assess the SEAES project at AUL 6.  

\subsection{Transformative and translational research identified by the needs of the user: Phase III AUL 7 project}\label{append:Carl}
\noindent {\bf C. J. Henney}

Global solar magnetic maps are the primary input driver for most coronal and solar wind models, however, the assembling of such maps is challenging since the solar photospheric and chromospheric magnetic fields are currently only recorded for less than half of the solar surface at any given time. With a limited view of the sun, and the rotational period of the Sun as observed from Earth is approximately 27 days, global maps of the magnetic field include old data, ranging from 15 days at mid-latitudes to 6 months at the poles. The primary goal of the ADAPT (Air Force Data Assimilative Photospheric Flux Transport) project \cite{Arge10,Arge11,Hick15} is to provide sequences of best estimates of the instantaneous global spatial distribution of the solar photospheric magnetic field as a function of time. Initiated in 2008 and driven by community user interests, the objective of the ADAPT project began by combining two projects,a photospheric magnetic flux transport model based on \citet{Word00} and rigorous data assimilation based on Kalman Filtering \cite{Hick15}, to produce global magnetic maps with realistic estimates of the uncertainty (satisfying milestones a-c of AUL 1).

An essential element during the ADAPT model development has been the vital feedback and collaboration with active users (satisfying milestones a-c in AUL 2) to assess the viability of the global maps within different scientific contexts (satisfying milestones a - d AUL 3). For example, the ADAPT global maps have been used with time-dependent MHD simulations of the inner heliosphere \cite{Merk2016}, new techniques for driving non-potential solar coronal magnetic field modeling \cite{Wein2016}, ensemble modeling of the large CME during July 2012 \cite{Cash2015}, scale-dependent data assimilation of solar photospheric magnetic fields \cite{Hick2016}, and empirically driven time-dependent modeling of the solar wind \cite{Link2016}. Another fundamental step during the project development was to integrate the ADAPT software within a prototyping environment (satisfying milestones a-b in AUL 4) and iterating on map quality, along with meta-data improvements, with various users (milestones a-b AUL 5). For example, the ADAPT model has been running autonomously at the National Solar Observatory for the past 5 years, generating public global magnetic maps for user validation (satisfying milestones a-c AUL 6). Integrating  and running ADAPT autonomously within a prototype system, identifying and managing challenges, integrating the components, and prototype the system in a simulated operational environment, provided the critical real-world testing and feedback needed to ready the ADAPT software to be installed and run on demand at NOAA-SWPC. The core functionality of the ADAPT model and this specific application is in the early stages milestones a-b of AUL 8, now released and installed at NOAA/SWPC to be validated in the context of driving WSA-Enlil, in collaboration with the CCMC. For more background on ADAPT, and access to real-time ADAPT global solar magnetic maps, see www.nso.edu/data/nisp-data/adapt-maps/.

While searching for full-disk integrated metric parameters to validate the timing and amplitude of far-side flux evolution and emergence within ADAPT maps, a significant new application branched off the ADAPT development (Figure \ref{fig_branch}), the SIFT (Solar Indices Forecasting Tool) empirical models. A preliminary viability study of full-disk integrated parameters for global map feedback led to the discovery that flux transport modeling can be utilized to predict the observed F$_{10.7}$ (i.e., solar radio flux at 10.7 cm) values \cite{Henn2012} and bands within the VUV (vacuum ultraviolet, between 0.1 and 175 nm, which includes the XUV, EUV, and FUV) solar irradiance \cite{Henn2015}. Solar F$_{10.7}$ and EUV are both key inputs to ionospheric and thermospheric models, and the ability to forecast these quantities more reliably allows for the possibility of advanced prediction of satellite drag and ionospheric structure, as proposed in Section \ref{Jeff}. After completing the basic research ][see ][]{Henn2015}, iterating with users on the quality (i.e., showed improvement compared to models utilized by users) and final forecast product format (AUL 1-5), the SIFT model for F$_{10.7}$ has been user validated and is operating autonomously in a prototype mode (satisfies requirements up to and including AUL 7) generating public predictions for users, along with providing real-time feedback on the ADAPT maps. The next step for the SIFT F$_{10.7}$ forecast model, to be assessed at AUL 8, is to validate the user application metrics were met within the time specifications. For more background on SIFT, and access to real-time SIFT forecasts, see www.nso.edu/data/nisp-data/sift-forecasts/.

\subsection{Example of a Phase III AUL 9 Project: Nowcast of keV electrons in the inner magnetosphere with IMPTAM - N.Yu. Ganushkina \label{append:NYG}}

The development of the Inner Magnetosphere Particle Transport and Acceleration Model (IMPTAM) was started as a tool to explain the observed features of ion dispersed structures in the inner magnetosphere seen at energy-time spectrograms from the CAMMICE/MICS instrument onboard the Polar spacecraft \cite{Ganushkina2000, Ganushkina2001}. One of the important results obtained from IMPTAM modeling was the ability of the model to reproduce the observed amount of ring current protons with energies $>$ 80 keV during a storm recovery phase \cite{Ganushkina2005, Ganushkina2006} by incorporating, in addition to the large-scale fields, transient fields associated with the dipolarization process in the magnetotail during substorm onset.  The name IMPTAM appeared, actually, later in the study of the dependence of the modeled ring current on the representations of magnetic and electric fields and boundary conditions used in simulations \cite{Ganushkina2012}. For this initial project, the model was used for purely scientific purposes, without any identification of potential users or specific applications. 

IMPTAM began the path towards an application when the project called SPACECAST was funded by the European Union Seventh Framework Programme (FP7/2007-2013) in 2011 (ended in February 2014). The main goal of this project was formulated as protecting space assets from high energy particles by developing European dynamic modeling and forecasting capabilities. The SPACECAST team consisted of leading experts from several EU countries providing their models for the radiation environment for further development and inter-coupling inside the project. At that stage, the identified users for IMPTAM were BAS, British Antarctic Survey, (Cambridge, UK) and ONERA, Office National D'\'Etudes et de Recherches Aerospatiales, (Toulouse, France). British Antarctic Survey had their BAS radiation belts model \cite{GlauertHorne2005} and ONERA had their Salammb\^o global radiation belt model \cite{BeutierBoscher1995}. Neither model included low energy electrons, the seed population of $<$ 100 keV, which is critically important for radiation belt dynamics. The first nowcast version of IMPTAM for $<$ 100 keV electrons \cite{Ganushkina2013, Ganushkina2014} running online in real time (http://fp7\-spacecast.eu/) was developed providing seed population for both BAS and Salammb\^o models \cite{Horne2013}. All milestones in the three AULs were passed in Phase I: basic scientific concepts and potential applications were identified at the beginning of the SPACECAST project (AUL 1); the users together with their requirements were identified (AUL 2); and IMPTAM was at the current state of the art \cite{Ganushkina2014} being able to reproduce the observed variations of keV electrons on the time scale of minutes. All the project's deliverables were successfully submitted in time, all the deadlines were met which were requirements from the European Commission. 

IMPTAM moved toward milestones in Phase II of the AUL framework, even during the SPACECAST project, since it went through development, testing, and validation. Milestones required for AUL 4 and AUL 5 were satisfied when IMPTAM was integrated into the functioning system of radiation belt models running at http://fp7-spacecast.eu/. Validation of the IMPTAM output has been ongoing since initial operation online in real time in February 2013 \cite{Ganushkina2015}, satisfying the milestones for AUL 6. In a sense, IMPTAM could already satisfy Phase-III milestones for this application since it has been operational in the user environment for a full year when the SPACECAST project ended. 

The IMPTAM project addressed a new application during the next project, SPACESTORM (http://www.spacestorm.eu/), funded by the European Union Seventh Framework Programme (FP7/2007-2013) in 2013 (ended in March 2017). The milestones in AULs 4--6 were satisfied and these efforts written up in the project's deliverables for the Commission. This application of IMPTAM fully entered into Phase III as the integrated system was implemented at the users' system. That is, IMPTAM was implemented and integrated into operational status. 

The SPACESTORM consortium consisted of five partners and the goal was to model severe space weather events and mitigate their effects on satellites by developing better mitigation guidelines, forecasting, and by experimental testing of new materials and methodologies to reduce vulnerability. During the SPACESTORM project, a ''real-world" user was identified which was the group of project participants from ONERA's DESP, Space Environment Department. The presence of rapidly-varying low energy ($<$200 keV) electrons causes surface charging effects on satellites, changes in the satellite potential and deg-radiation of satellite surface materials. Therefore, the unique value of IMPTAM's ability to model the variations of keV electron fluxes at any satellite orbit in the inner magnetosphere was of the exceptional interest for users from ONERA. They identified a current need to determine the risks that extreme events present to critical spacecrafts in GEO and MEO (geosynchronous and medium Earth orbit, respectively). The special software called Spacecraft Plasma Interaction Software (SPIS) has been developed at ONERA under ESA (European Space Agency) and CNES (Centre National D'\'Etudes Spatiales, the French government space agency) funding. SPIS is used to assess surface charging levels of spacecraft immersed in severe GEO and MEO environments. The requirements set for IMPTAM were to provide locations and magnitudes of worst-case electron fluxes ($<$ 100 keV) at MEO by validating IMPTAM at GEO based on the database of surface charging events observed at LANL spacecraft \cite{MateoVelez2018}. This procedure follows all three AULs for Phase III:  AUL 7 with application prototype in which the type and specifics of application was determined, AUL 8 with validation in relevant ''real world" environment with the observed surface charging events, and AUL 9 with the on-demand use of IMPTAM by ONERA's SPIS software to compute surface charging for any event of interest. 

At present, IMPTAM is part of the on-going project PROGRESS (ssg.group.shef.ac.uk$/$progress$/$html) funded by the European Union's Horizon 2020 research and innovation programme (ends in July 2018). The overall aim of the PROGRESS project is to develop an accurate and reliable forecast of space weather hazards. In this project, IMPTAM is undergoing transformations to operate as a predictive tool (imptam.fmi.fi, https:$//$ssg.group.shef.ac.uk$/$progress$/$html$/$imptam\_results.phtml), not only as a near real-time tool which it has been so far. To be a predictive tool, IMPTAM required the foretasted solar wind and IMF parameters and geomagnetic indices to drive it. IMPTAM can be considered predictive when reliable forecasts for its driving parameters become available within the PROGRESS project.

\pagebreak
\section{The AUL Checklists}\label{checklist}
We have provided easy to use checklists for readers who want to use the AUL framework. 
\begin{figure}[ht]
\centering\includegraphics[width=0.9\linewidth]{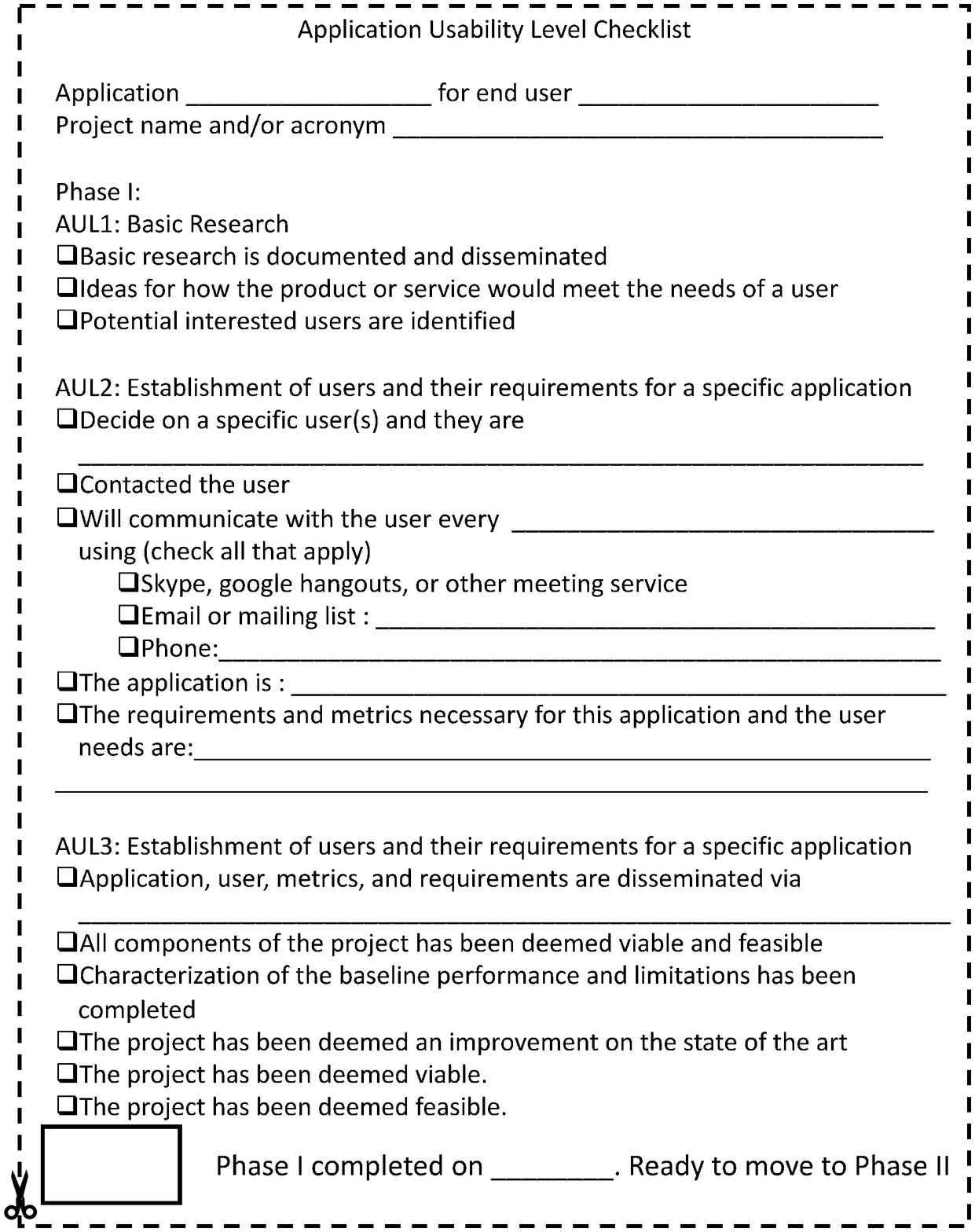}
\caption{Application Usability Level (AUL) Checklist for Phase 1.}\label{Fig_AUL_check1}
\end{figure}

\pagebreak

\begin{figure}[ht]
\centering\includegraphics[width=0.9\linewidth]{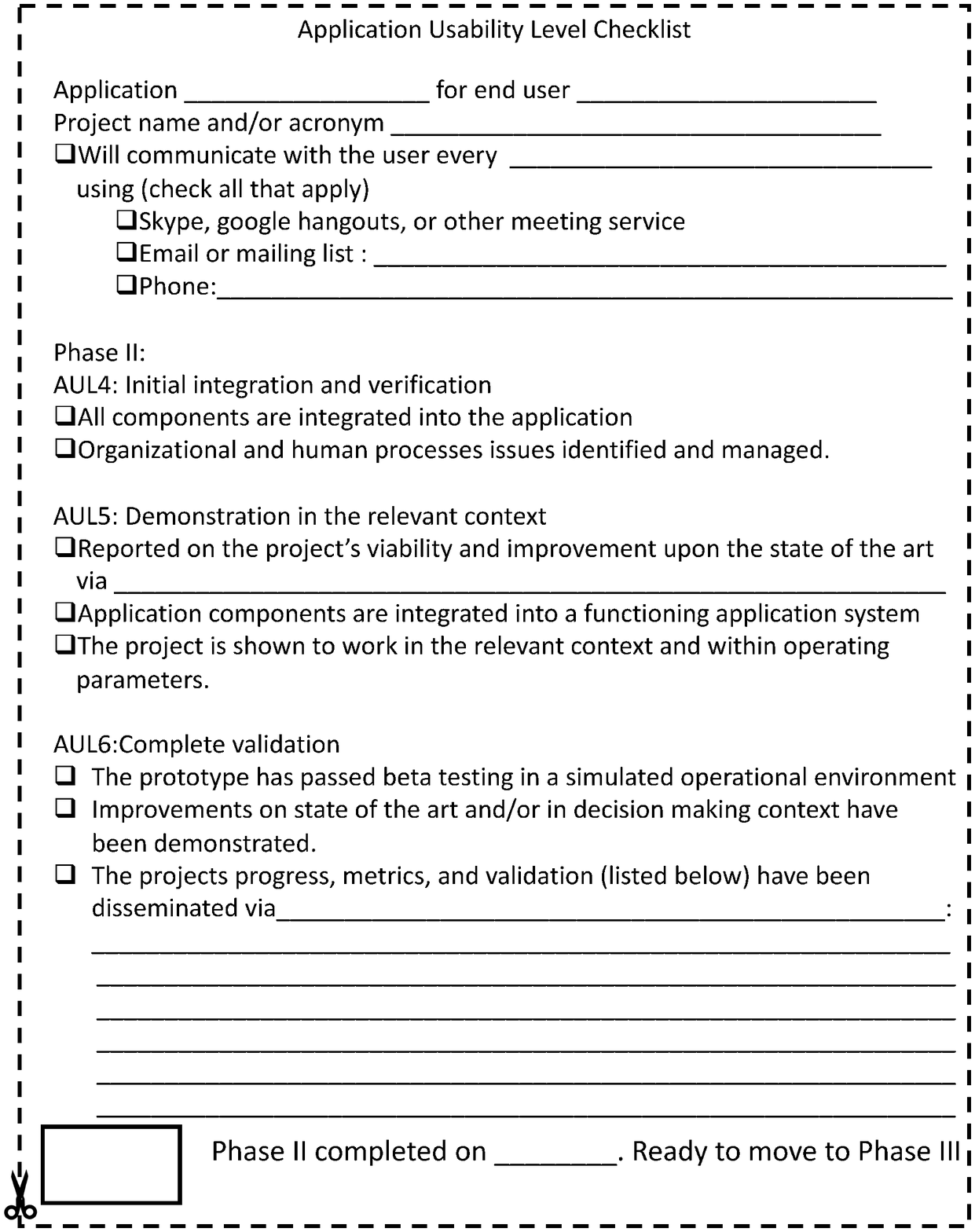}
\caption{Application Usability Level (AUL) Checklist for Phase 2. }\label{Fig_AUL_check2}
\end{figure}

\pagebreak 
\begin{figure}[ht]
\centering\includegraphics[width=0.9\linewidth]{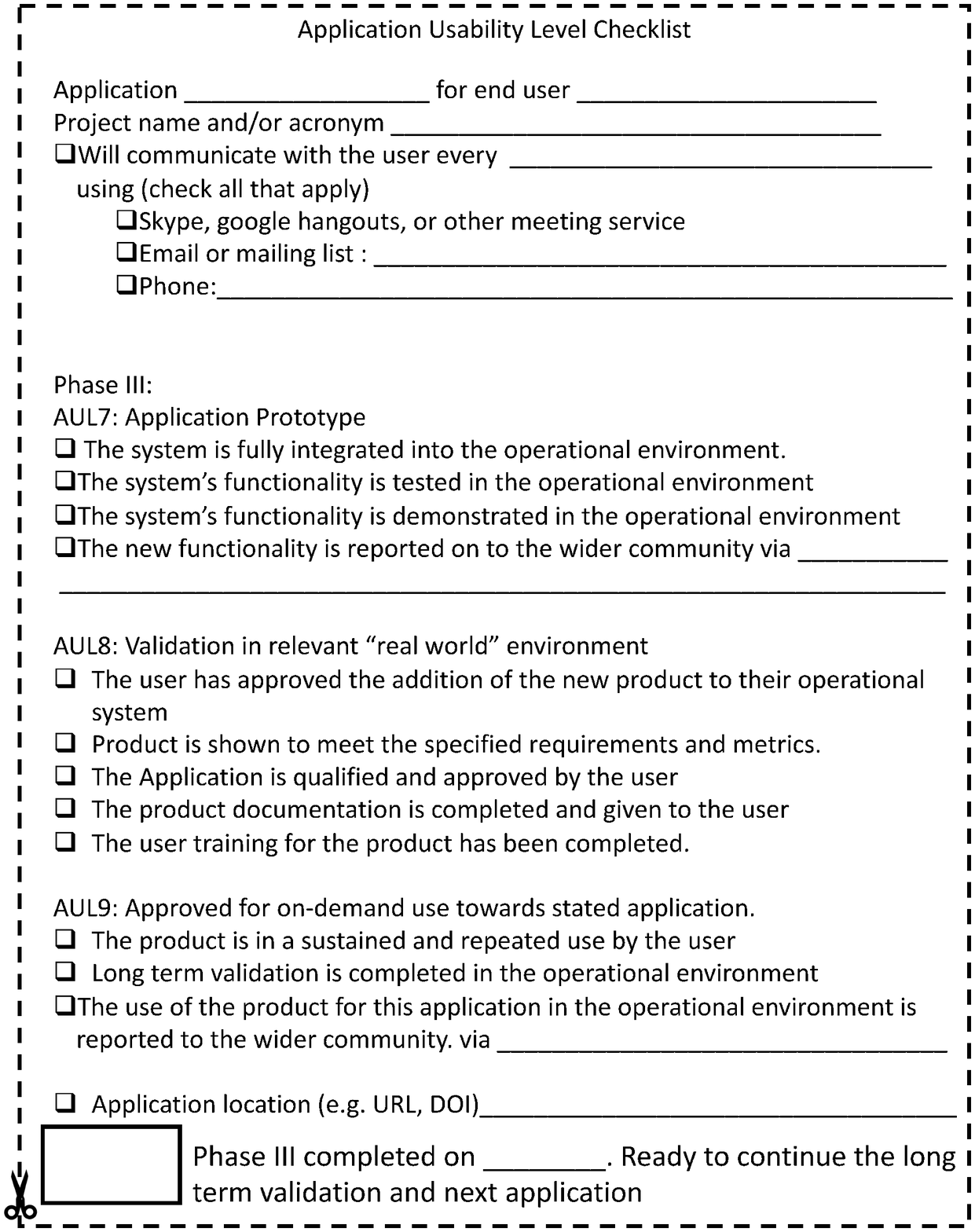}
\caption{Application Usability Level (AUL) Checklist for Phase 3. }\label{Fig_AUL_check3}
\end{figure}

\pagebreak
\begin{acknowledgements}
This work has been completed as part of the Assessment of Understanding and Quantifying Progress working group which is part of the International Forum for Space Weather Capabilities Assessment. A large thanks goes to the International Forum for Space Weather Capabilities Assessment, CCMC, Masha Kuznetsova, and Lawrence Friedl for the wonderful discussions during the  workshop and since which resulted in this paper. The authors would like to thank overleaf which aided in our ability to collaborate among a large number of active authors. A.J.H. would like to thank all of the co-authors for their efforts and patience with putting together such a large collaboration, especially such an international one where co-authors were at times asked to stay up very late or get up way too early. All authors provided help throughout the entire paper but A.J.H. would like to point out where individual efforts were focused and greatly appreciated. Specifically, A.J.H. would like to thank, B.A.C, J.K, R.M.M, C.C., C.J.H., A.G.B., T.G., N.Y.G., and M.T. for their extensive work on providing examples for the paper and being Guinea Pigs for applying the AUL framework to their past and ongoing projects; M.W.L, B.M.W, K.D.L, S.F., M.M.B, and S.K.M., for their help writing and refining the intro and discussion sections and especially with help defining all the vocabulary; D.T.W. and A.G.B. for going above and beyond with help to state all of the communities ideas throughout the paper as succinctly as possible; A.P, and B.J.T for bringing their past experience using and working with ARLs to help refine the new AUL levels and making them more applicable and useful for the space weather community; K.G.-S. for helping with the entire paper and bringing a focus to past and present validation efforts from around the Heliophysics community; S.B., S.A.M, and J.P.M.  for bringing the focus of the space weather user community to the paper; And last but definitely not least, A.C.K. - well, every first author needs a fantastic second author to help coordinate meeting efforts, providing an example project, and help outlining a paper as well as a cheerleader when things seem daunting. Along with all the co-authors, we would like to thank the larger Space Physics/Space Weather/Heliophysics community which has provided feedback and comments at many telecoms, conferences, and workshops. Along with the wider community, we'd like to extend a thank you to the reviewers for helping to make this paper more accessible, and encouraging us to provide tools and an infrastructure to help with easy implementation by the community. We hope through this large and active collaboration from across the community, we have developed and provided a new useful tool for applied and pure targeted space weather and space physics research.

The proposals that in part funded R. M. M's Ph.D. research which is used as an example in section 4.2 was the NSF Fellowship award DGE 1144083, NSF grant AGS-1025089, NASA grant NNX13AD64G, and the Vela Fellowship at the Los Alamos National Labs Space Weather Summer School. Portions of this research were carried out at the Jet Propulsion Laboratory, California Institute of Technology, under a contract with the National Aeronautics and Space Administration. RMM is currently supported by the NASA Living With a Star Jack Eddy Postdoctoral Fellowship Program, administered by the University Corporation for Atmospheric Research and coordinated through the Cooperative Programs for the Advancement of Earth System Science (CPAESS). S.A.M. is supported by the Irish Research Council Postdoctoral Fellowship Programme and the Air Force Office of Scientific Research award number FA9550-17-1-039. The projects leading to the results presented by N. Yu. G. have received funding from the European Union Seventh Framework Programme (FP7/2007-2013) under grant agreement No 606716 SPACESTORM and from the European Union's Horizon 2020 research and innovation program under grant agreement No 637302 PROGRESS. Work of M.W.L. and N. Yu. G. in the US was conducted under NASA grants NNX17AI48G, NNX17AB87G, and 80NSSC17K0015, and NSF grant 1663770. B.A.C. and M.T.'s input was funded by an Australian Research Council Linkage Project grant (LP160100561). SKM acknowledges support from the US Department of Energy's Laboratory Directed Research and Development program (grant number 20170047DR). A.G.B.'s work is supported by the Office of Naval Research. ACK acknowledges support from NASA grant NNX16AG78G and NSF grant AGS-1552321. AJH's work is supported in part by The Aerospace Corporation. This work did not provide any new datasets. 

The editors are grateful to Daniel Heynderickx and an anonymous referee who evaluated an earlier version of this paper and whose reviews were kindly made available to the editors of JSWSC.
\end{acknowledgements}


\bibliography{refs}
   

\end{document}